%% file: LoI_MuEDM.tex
\documentclass[aps,superscriptaddress,showkeys,a4paper]{revtex4}

\usepackage{aas_macros}
\usepackage{amsmath}
\usepackage{amssymb}
\usepackage{bm}
\usepackage[dvipsnames,usenames]{color}
\usepackage[pdftex]{graphicx,epsfig}
\usepackage[caption=false]{subfig}
\usepackage{bm}
\usepackage[tight,nice]{units}
\usepackage{siunitx}
\DeclareSIUnit\micron{\micro\metre}
\usepackage[breaklinks=true,hidelinks]{hyperref}

\graphicspath{{./Images/}}

\input{CommandsAndShortcuts.tex}

\begin{document}

\title{Search for a muon EDM using the frozen-spin technique}

\author{A.~Adelmann}
\affiliation{ETH Zürich, 8093 Zürich, Switzerland}
\affiliation{Paul Scherrer Institut, 5232 Villigen PSI,
Switzerland} 
\author{M.~Backhaus}
\affiliation{ETH Zürich, 8093 Zürich, Switzerland}
\author{C.~Chavez~Barajas}
\affiliation{University of Liverpool, Liverpool, UK}
\author{N.~Berger}
\affiliation{PRISMA$^{+}$ Cluster of Excellence and Institute of Nuclear Physics, Johannes Gutenberg University Mainz, Mainz, Germany}
\author{T.~Bowcock}
\affiliation{University of Liverpool, Liverpool, UK}
\author{C.~Calzolaio}
\affiliation{Paul Scherrer Institut, 5232 Villigen PSI,
Switzerland}
\author{G.~Cavoto}
\affiliation{Sapienza Universit\`a di Roma, Dip.\ di Fisica, P.le A.\ Moro 2, 00185 Roma, Italy}
\affiliation{Istituto Nazionale di Fisica Nucleare, Sez.\ di Roma, P.le A.\ Moro 2, 00185 Roma, Italy}
\author{R.~Chislett}
\affiliation{University College London, London, UK}
\author{A.~Crivellin}
\affiliation{Paul Scherrer Institut, 5232 Villigen PSI,
Switzerland}
\affiliation{CERN,  1211 Geneva, Switzerland}
\affiliation{University of Zürich, Zürich,
Switzerland}
\author{M.~Daum}
\affiliation{Paul Scherrer Institut, 5232 Villigen PSI,
Switzerland}
\author{M.~Fertl}
\affiliation{PRISMA$^{+}$ Cluster of Excellence and Institute of Physics, Johannes Gutenberg University Mainz, Mainz, Germany}
\author{M.~Giovannozzi}
\affiliation{CERN,  1211 Geneva, Switzerland}
\author{G.~Hesketh}
\affiliation{University College London, London, UK}
\author{M.~Hildebrandt}
\affiliation{Paul Scherrer Institut, 5232 Villigen PSI,
Switzerland} 
\author{I.~Keshelashvili}
\affiliation{Institut f\"ur Kernphysik, Forschungszentrum J\"ulich, J\"ulich, Germany}
\author{A.~Keshavarzi}
\affiliation{Department of Physics and Astronomy, University of Manchester, Manchester, UK}
\author{K.S.~Khaw}
\affiliation{Tsung-Dao Lee Institute, Shanghai Jiao Tong University, Shanghai, China}
\affiliation{School of Physics and Astronomy, Shanghai Jiao Tong University, Shanghai, China}
\author{K.~Kirch}
\affiliation{ETH Zürich, 8093 Zürich, Switzerland}
\affiliation{Paul Scherrer Institut, 5232 Villigen PSI,
Switzerland}
\author{A.~Kozlinskiy}
\affiliation{PRISMA$^{+}$ Cluster of Excellence and Institute of Nuclear Physics, Johannes Gutenberg University Mainz, Mainz, Germany}
\author{A.~Knecht}
\affiliation{Paul Scherrer Institut, 5232 Villigen PSI,
Switzerland} 
\author{M.~Lancaster}
\affiliation{Department of Physics and Astronomy, University of Manchester, Manchester, UK}
\author{B.~M\"arkisch}
\affiliation{Physik-Department, Technische Universit\"at M\"unchen, Garching, Germany }
\author{F.~Meier~Aeschbacher}
\affiliation{Paul Scherrer Institut, 5232 Villigen PSI, Switzerland}
\author{F.~M\'eot}
\affiliation{Brookhaven National Laboratory, USA}
\author{A.~Nass}
\affiliation{Institut f\"ur Kernphysik, Forschungszentrum J\"ulich, J\"ulich, Germany}
\author{A.~Papa}
\affiliation{Paul Scherrer Institut, 5232 Villigen PSI,
Switzerland}
\affiliation{University of Pisa and INFN, Pisa, Italy}
\author{J.~Pretz}
\affiliation{Institut f\"ur Kernphysik, Forschungszentrum J\"ulich, J\"ulich, Germany}
\affiliation{RWTH Aachen, Aachen, Germany}
\author{J.~Price}
\affiliation{University of Liverpool, Liverpool, UK}
\author{F.~Rathmann}
\affiliation{Institut f\"ur Kernphysik, Forschungszentrum J\"ulich, J\"ulich, Germany}
\author{F.~Renga}
\affiliation{Istituto Nazionale di Fisica Nucleare, Sez.\ di Roma, P.le A.\ Moro 2, 00185 Roma, Italy}
\author{M.~Sakurai}
\affiliation{ETH Zürich, 8093 Zürich, Switzerland}
\author{P.~Schmidt-Wellenburg}
\email{philipp.schmidt-wellenburg@psi.ch} \affiliation{Paul Scherrer Institut, 5232 Villigen PSI,
Switzerland} 
\author{A.~Sch\"oning}
\affiliation{Physics Institute, Heidelberg University, Heidelberg, Germany}
\author{M.~Schott}
 \affiliation{PRISMA$^{+}$ Cluster of Excellence and Institute of Nuclear Physics, Johannes Gutenberg University Mainz, Mainz, Germany}
\author{C.~Voena}
\affiliation{Istituto Nazionale di Fisica Nucleare, Sez.\ di Roma, P.le A.\ Moro 2, 00185 Roma, Italy}
\author{J.~Vossebeld}
\affiliation{University of Liverpool, Liverpool, UK}
\author{F.~Wauters}
\affiliation{PRISMA$^{+}$ Cluster of Excellence and Institute of Nuclear Physics, Johannes Gutenberg University Mainz, Mainz, Germany}
\author{P.~Winter}
\affiliation{Argonne National Laboratory, Lemont, USA}

\begin{abstract}
This letter of intent proposes an experiment to search for an electric dipole moment of the muon based on the frozen-spin technique. We intend to exploit the high electric field, $E=\SI{1}{GV/m}$, experienced in the rest frame of the muon with a momentum of $p=\SI{125}{MeV/}c$ when passing through a large magnetic field of $|\vec{B}|=\SI{3}{T}$. Current muon fluxes at the $\mu$E1 beam line permit an improved search with a sensitivity of $\sigma(d_\mu)\leq\SI{6e-23}{\ecm}$, about three orders of magnitude more sensitivity than for the current upper limit of $|d_\mu|\leq\SI{1.8e-19}{\ecm}$\,(C.L. 95\%). With the advent of the new high intensity muon beam, HIMB, and the cold muon source, muCool, at PSI the sensitivity of the search could be further improved by tailoring a re-acceleration scheme to match the experiments injection phase space.
While a null result would set a significantly improved upper limit on an otherwise un-constrained Wilson coefficient, the discovery of a muon EDM would corroborate the existence of physics beyond the Standard Model. 
\end{abstract}

\maketitle
\newpage
\section{Summary}
This letter of intent proposes a search for the electric dipole moment (EDM) of the muon. We plan to design, mount, and operate an experiment to search for an EDM of the muon (muEDM), using the frozen-spin technique~\cite{Farley2004PRL} at the High Intensity Proton Accelerator (HIPA) of the Paul Scherrer Institute (PSI). 

Particle EDMs are generally considered as excellent probes of physics Beyond the Standard Model~(BSM)~\cite{Raidal2008}, indicating the violation of the combined symmetry of charge and parity~(CPV). Indeed CPV is one of three necessary conditions
%, first described by Sakharov in the 1970's\,\cite{Sakharov1967}, 
to explain the creation of a matter-dominated Universe from an initially symmetric condition~\cite{Sakharov1967}. The observed Baryon Asymmetry of the Universe (BAU)~\cite{Komatsu2011} cannot be explained by the otherwise successful Standard Model (SM) of particle physics~\cite{Morrissey2012NJP} as, among others, the existing CPV in the weak sector of the SM is insufficient.
% Searches for an EDM of the electron~\cite{Baron2014Science}, the neutron~\cite{Pendlebury2015PRD}, and of mercury~\cite{Graner2016PRL}  set impressive limits and are important benchmarks for theories beyond the SM (BSM)~\cite{Chupp2015PR,Engel2013PPNP}. 

The EDM limit of the muon $d_\mu \leq \unit[\pow{1.8}{-19}]{\ecm}$~(95\% C.L.)\,\cite{Bennett2009PRD} is the only EDM of a fundamental particle probed directly on the bare particle. Assuming simple scaling in mass, by the ratio ($m_{\mu}/m_e$ and lepton universality the electron EDM, $d_e \leq\SI{1.1e-29}{\ecm}$~(95\% C.L.)~\cite{Andreev2018}, measured using thorium monoxide~(ThO) molecules provides a much tighter indirect limit, assuming the electron is the only source of CPV\@. As we will argue in \autoref{sec:Motivation}, current $B$-meson decay anomalies at LHC~\cite{Altmannshofer2017EPJC} and the persistent $3.7\,\sigma$ discrepancy of the muon $g-2$ motivate that new physics may have flavor structure beyond the paradigm of Minimal Flavor Violation (MFV), removing possible constraint on the muon EDM from other lepton EDM searches. 

\begin{figure}[h]
	\centering
			% 1
			\subfloat[]{
					\includegraphics[width=0.48\columnwidth]{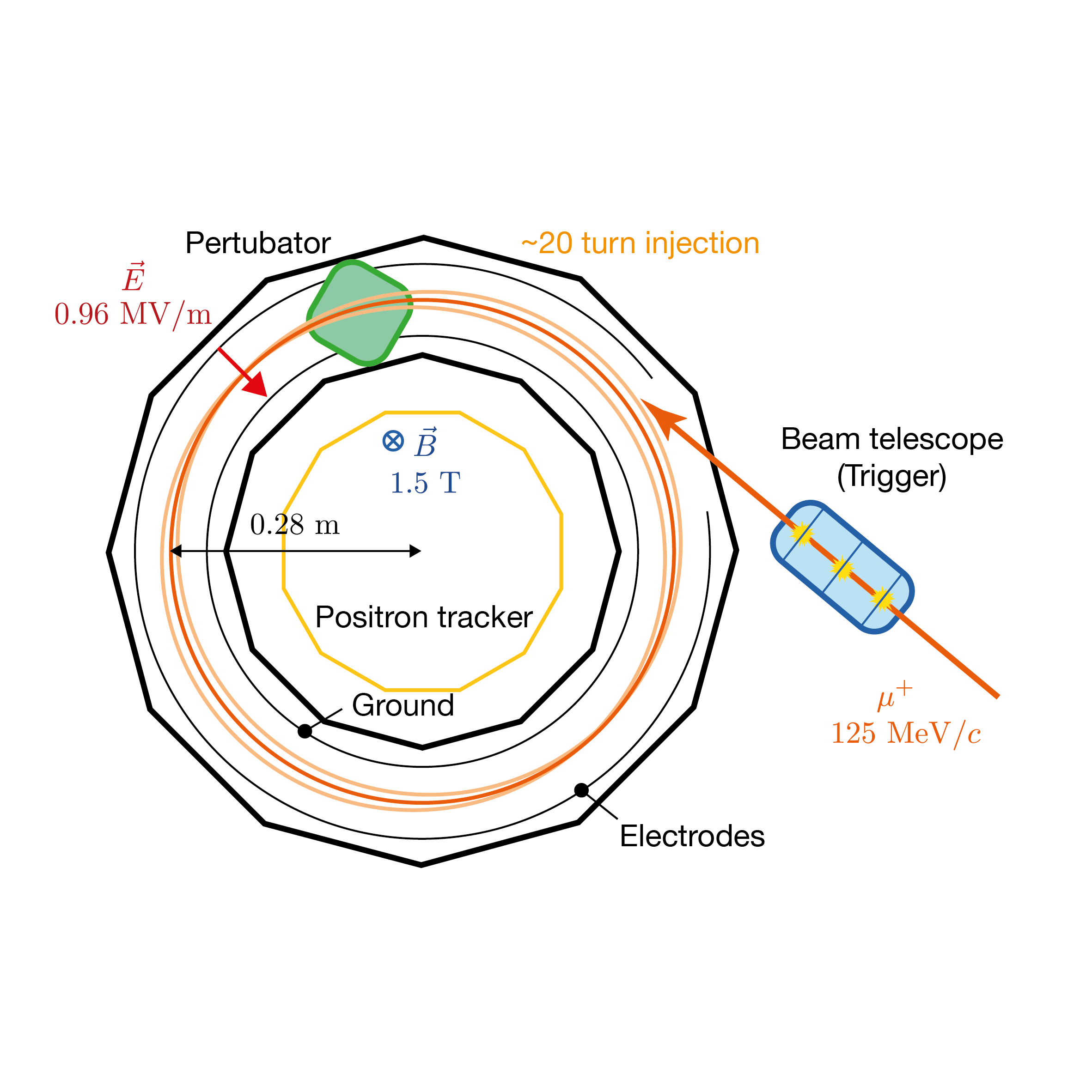}
					\label{fig:StorageRingMuEDM}
			}	
			\hfill
			\subfloat[]{
					\includegraphics[width=0.48\columnwidth]{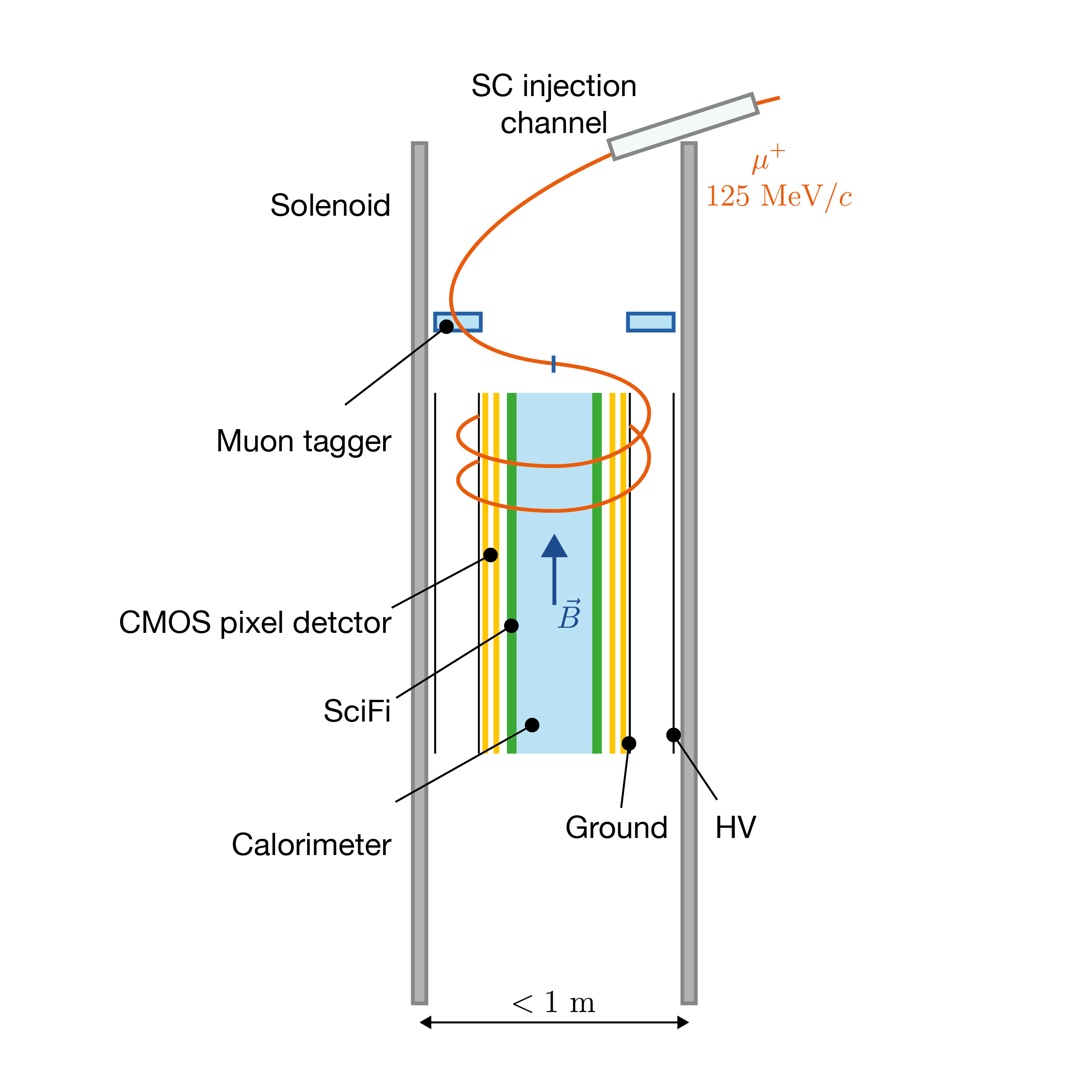}
			}
		\caption{Sketch of storage ring with lateral injection~(a), and the helix muEDM search~(b) using a vertical injection into a uniform solenoid field.}%
		\label{fig:ComparisonStorageVsHelix}%
\end{figure}

We therefore propose a dedicated experiment, which permits the search for the muEDM with a sensitivity of about \SI{6e-23}{\ecm} per year of data-taking. The baseline concept plans to use muons with a momentum of $p=\SI{125}{MeV/}c$ ($\beta = 0.77$) and an average polarization of 90\% from the $\mu$E1 beam line at PSI with a particle flux of up to $\unit[\pow{2}{8}]{\mu^+/s}$. Two concepts are currently under evaluation and discussed in this letter of intent. In \autoref{sec:storagering} we discuss a storage ring sketched in  \autoref{fig:ComparisonStorageVsHelix}a, with a magnetic field of $|\vec{B}|=$\SI{1.5}{T} where muons are injected laterally, similar as described in~\cite{Adelmann2010JPG}.
The second concept, the helix muEDM in \autoref{fig:ComparisonStorageVsHelix}b, is based on the idea of a vertical injection into a \SI{3}{T} B-field similar as proposed by the J-PARC~$(g-2)$ group\,\cite{Iinuma2016NIMA} and is discussed in \autoref{sec:3DInjection}.
%with a short field-free section 

In both cases a nested electrode system provides a radial electric field $E_{\rm f}$ for the frozen-spin technique, discussed in~\autoref{sec:frozenspin}. Positive muons will be stored one at a time on a stable orbit inside the frozen-spin region. The muon sees a large electric field $\vec{E} \approx \gamma\vec{\beta}\times\vec{B}$, about $\SI{1}{GV/m}$ for $|\vec{B}|=\SI{3}{T}$, which leads to a precession of the spin in the presence of a muEDM, while the oscillation from the anomalous magnetic moment is suppressed. The muons will decay after an average lifetime of $\gamma\tau_\mu=\SI{3.4}{\micro\second}$ in the lab system into a positron and two neutrinos. Due to the parity violating decay, the positron is preferentially emitted along the spin of the muon, with an average asymmetry of $A=0.3$. By detecting the vertical asymmetry of positrons ejected upwards or downwards with a tracker placed inside the helix/orbit the build-up in time of an asymmetry due to an EDM will be measured.

\newpage
\section{Motivation}
\label{sec:Motivation}
A non-zero EDM of a fundamental particle violates time-reversal symmetry, and by invoking the CPT-theorem of quantum field theories~\cite{Luders1954}, also the combined symmetry of charge conjugation and parity inversion (CP). Many BSM theories have new complex parameters which are sources of CP violation as these parameters are naturally expected to have a generic phase of order one. In fact, the only complex parameter within the SM (disregarding the vanishingly small QCD theta term), the phase of the Cabibbo Kobayashi Maskawa (CKM) matrix~\cite{Kobayashi1973}, is close to maximal~\cite{Ciuchini2000,Hocker:2001xe}. Furthermore, CP~violation is also one of three necessary conditions to explain the observed BAU~\cite{Sakharov1967}. However, even though the CKM phase is close to maximal, CP~violation within the SM is by far not sufficient to explain the observed BAU~\cite{Cohen:1993nk,Gavela:1993ts,Huet:1994jb,Gavela:1994ds,Gavela:1994dt,Riotto:1999yt}. This strongly motivates theories with additional complex parameters as extensions of the SM, providing additional sources of CP violation. Clearly, such sources of CP violation are expected to generate at some level non-vanishing electric dipole moments of fundamental particles, which can significantly exceed the tiny values within the SM~\cite{Engel2013PPNP}.

\begin{figure}%
\includegraphics[width=0.7\columnwidth]{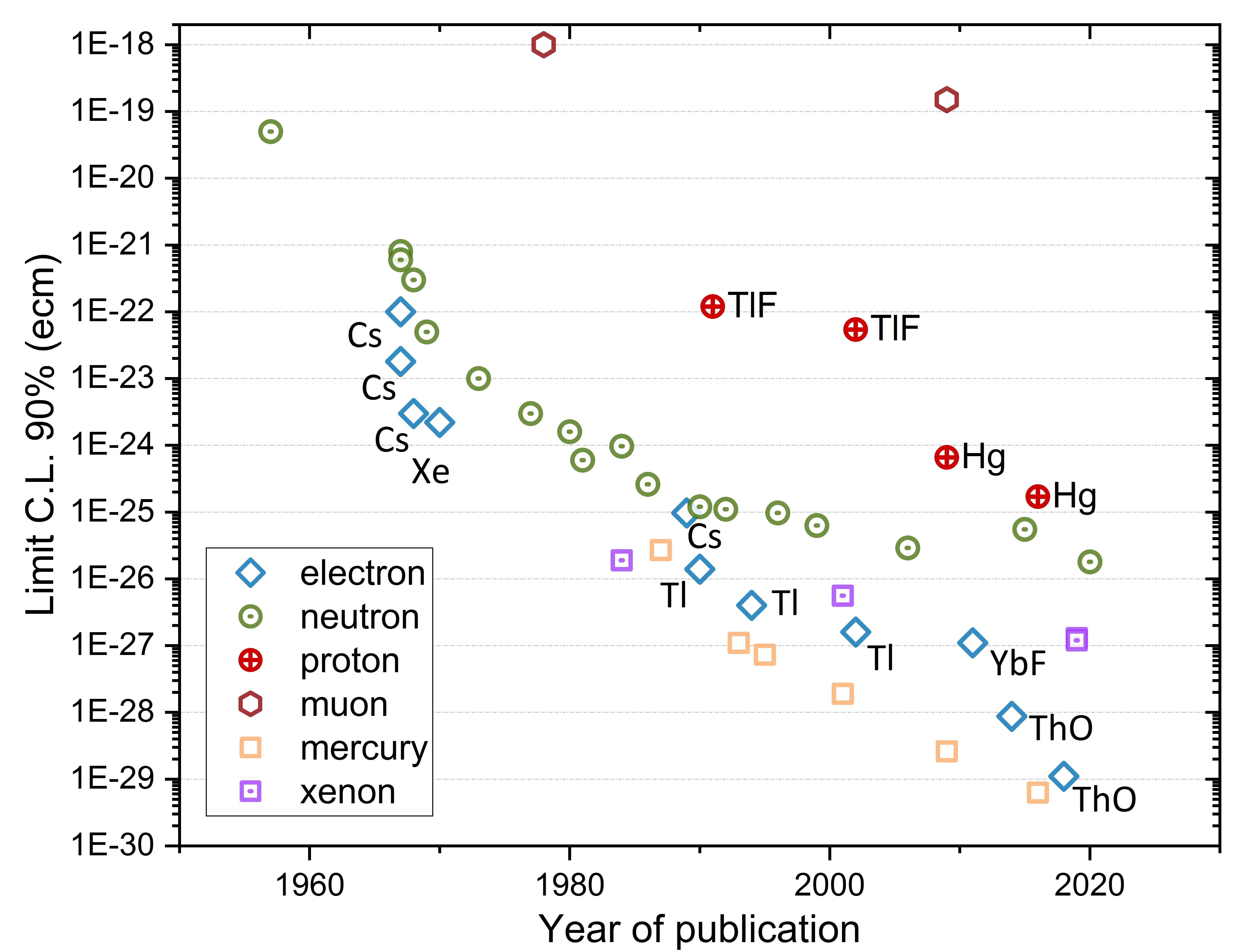}%
\caption{Historical overview of EDM limits (90\% C.L.). The labels in the plot next to the date (Cs, Tl, TlF, ThO, Xe, and YbF) refer to the measured system from which the limit was derived. So far, all EDM measurements were in agreement with a null result and were therefore interpreted as upper limits. }%
\label{fig:EDMOverview}%
\end{figure}

Therefore, many experiments searching for non-vanishing electric dipole moments have been performed over the last decades, as summarized in Figure~\ref{fig:EDMOverview}, and the current status can be found in \cite{Chupp2019RMP}. As we can see, the limits on the muon EDM are particularly weak compared to the other constraints. Therefore, a search for a permanent EDM of the muon gives access to one of the least tested areas of the SM of particle physics and is hence an important piece of this comprehensive and complementary experimental strategy to unveil BSM physics~\cite{ESPP2019}. 

One reason why in the past the focus of EDM searches was obviously not on the muon EDM is that the impressive limits on the electron EDM from measurements using atoms or molecules, e.g.\ thorium oxide molecules $d_e<\SI{1.1e-29}{\ecm}$~\cite{Andreev2018}, were commonly rescaled, assuming MFV~\cite{Chivukula:1987fw,Hall:1990ac,Buras:2000dm,DAmbrosio:2002vsn} (by the ratio $m_\mu/m_e$) resulting in $d_\mu <\SI{1.6e-27}{\ecm}$. However, MFV is, to some extent, an ad hoc symmetry invented to allow light particle spectra, in particular within the MSSM where this reduces the degree of fine-tuning in the Higgs sector while respecting at the same time flavor constraints. 
Since the LHC did not discover any new particles directly~\cite{Butler:2017afk,Masetti:2018btj} the whole concept of naturalness is challenged. Furthermore, LHCb, Belle and BaBar discovered significant tensions in semi-leptonic $B$ decays~\cite{Aaij:2014pli,Aaij:2014ora,Aaij:2015esa,Aaij:2015oid,Khachatryan:2015isa,ATLAS:2017dlm,CMS:2017ivg,Aaij:2017vbb} implying a  $5\,\sigma$ level discrepancy when analyzed together~\cite{Amhis2017EPJC,Altmannshofer2017PRD,Capdevila2018JHEP}.
These remarkable hints for new physics point towards the violation of Lepton Flavor Universality (LFU) and are therefore not compatible with MFV in the lepton sector~\cite{Cirigliano:2005ck}.

Furthermore, there is the longstanding 3.7$\,\sigma$ tension between the measured value of the anomalous magnetic moment (AMM) of the muon~\cite{Bennett2006PRD} and its SM prediction~\cite{Aoyama:2020ynm}. The AMM is directly related to the EDM since the former measures the real part of the same Wilson coefficient whose imaginary part gives rise to the non-vanishing EDM. While the measurement of the AMM of the muon is by itself consistent with the assumption of MFV, in general any TeV scale explanation of the AMM of the muon requires a chirally enhanced effect that automatically provides an a priori free phase. 
For example, the $B$ anomalies motivate the introduction of leptoquarks, which can account not only for them, but at the same time for the AMM of the muon~\cite{Crivellin:2019dwb} via a $m_t/m_\mu$ enhanced effect~\cite{Djouadi:1989md,ColuccioLeskow:2016dox} whose phase is completely unconstrained. 
%Furthermore, the recently observed tension in the AMM of electron~\cite{Parker2018} has opposite sign than the one in the muon sector and is thus not compatible with the MFV hypothesis~\cite{Crivellin:2018qmi}. 

\begin{figure}%
\includegraphics[width=0.5\columnwidth]{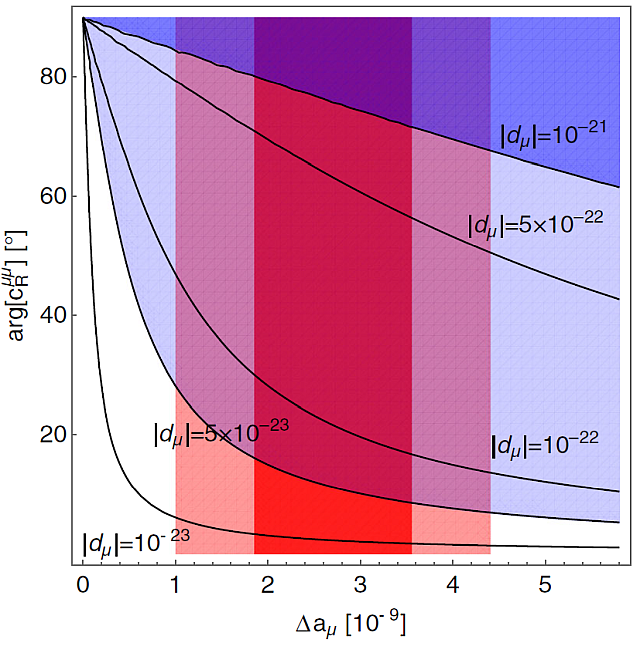}%
\caption{Contours of $d_\mu$ as a function of the anomalous momentum $\Delta a_\mu$ and the phase of the associated Wilson coefficient~\cite{Crivellin2019}.}
\label{fig:muEDMvsamu}%
\end{figure}

Therefore, it is well-motivated that New Physics~(NP) has a flavor structure beyond MFV\@. A notion often contested on grounds of naturalness arguments. However, note that in the limit of vanishing neutrino masses, which is an excellent approximation taking into account their smallness, lepton flavor is conserved. Thus it possible to completely disentangle the muon from the electron EDM via a symmetry, meaning that no fine-tuning is necessary. This could for example be achieved via a $L_\mu-L_\tau$ symmetry~\cite{He:1990pn,Foot:1990mn,He:1991qd} which can naturally give rise to the observed PMNS matrix~\cite{Binetruy:1996cs,Bell:2000vh,Choubey:2004hn}, and, even after its breaking, protects the electron EDM and AMM from NP effects~\cite{Altmannshofer:2016oaq}.
Also from an EFT point of view~\cite{Pruna2017}, it is clear, that the muon EDM can be large and that a measurement of it is in practice the only way of determining the imaginary part of the associated Wilson coefficient. 
In summary, this clearly demonstrates that a more sensitive measurement of the muon EDM has the potential to discover CP violation and further corroborates the existing hints for the violation of LFU~\cite{Crivellin2019}. This can be clearly seen from Figure~\ref{fig:muEDMvsamu} which shows the potential reach for the complex phase of the Wilson coefficients of a future muon EDM search at PSI\@. 
In fact, a discovery of a non-vanishing muon EDM would consolidate the existence of physics beyond the SM and lead to a paradigm shift in our understanding of nature.

\section{Experimental search for a muon EDM}
\subsection{Spin motion of muons in electric and magnetic field in the presence of an EDM}
The spin dynamics of a muon at rest in a magnetic field $\vec{B}$ is described by $\diff{\vec{s}}/\diff{t}=\vec{\mu}\times\vec{B}=\vec{\omega}_{\rm L}\times\vec{s}$ where
$\vec{\mu}=ge/(2m)\vec{s}$ is the magnetic dipole moment with $|\vec{s}|=\hbar/2$ and
%$\vec{\omega}_{\rm L}=2\vec{\mu}\!\cdot\!\vec{B}/\hbar$
$\vec{\omega}_{\rm L}=-2\mu\vec{B}/\hbar$ the Lamor precession frequency. Similarly, a hypothetical electric dipole moment $\vec{d}=\eta e/(2mc)\vec{s}$ results in a spin precession of the muon $\vec{\omega}_d=-2d\vec{E}/\hbar$ in an electric field $\vec{E}$.\newline 
The first search for a muEDM resulted in an upper limit of $\unit[\pow{2.9}{-15}]{\ecm}$~(95\% C.L.)\,\cite{Berley1958PRL,Berley1958PRL_Err} and was published in 1958. Half a century later, the current best upper limit of $d_\mu <\unit[\pow{1.8}{-19}]{\ecm}$ ~(95\% C.L.)\,\cite{Bennett2009PRD} was deduced using the spin precession data from the $(g-2)$ storage ring experiment E821 at BNL~\cite{Bennett2006PRD}. \\

For the further discussion of the spin dynamics of a moving muon with momentum $\vec{p}$, 
%$\vec{\beta}=\vec{p}/E$
$\vec{\beta}=\vec{v}/c$ and $\gamma=(1-\beta^2)^{-1/2}$ in magnetic, $\vec{B}$, and electric, $\vec{E}$, fields it is useful to change to the unit polarization three vector $\vec{\Pi}=\vec{s}/\left|\vec{s}\right|$. Then the change in polarization with time is given by

\begin{equation}
	\Diff{\vec{\Pi}}{t} = \vec{\Omega}_0\times\vec{\Pi},
\label{eq:polarizationDynamics}
\end{equation}
where 
\begin{equation}
	\vec{\Omega}_0 = -\frac{e}{m\gamma}\left[\left(1+\gamma a\right)\vec{B}-\frac{a\gamma^2}{\left(\gamma+1\right)}\left(\vec{\beta}\cdot\vec{B}\right)\vec{\beta}-\gamma\left(a+\frac{1}{\gamma+1}\right)\frac{\vec{\beta}\times\vec{E}}{c}\right]
\label{eq:ThomasPrecession}
\end{equation}
is the Thomas precession~\cite{Thomas1927}, when replacing the anomalous moment of the muon $a$~\cite{Bennett2006PRD} with $(g-2)/2$ and the parameter $\lambda$ in~\cite{Thomas1927} by $ge/(2mc)$.

In the case that no electric field is applied parallel to the momentum, the acceleration of the muon is purely transverse to its motion

\begin{equation}
	\Diff{\vec{\beta}}{t} = \frac{e}{\gamma mc}\left(\vec{E}+\vec{\beta}c\times\vec{B}\right),
\label{eq:LorentzEquation}
\end{equation}
which is equivalent to 
\begin{equation}
	\Diff{\vec{\beta}}{t} = \vec{\Omega}_c\times\vec{\beta},
\label{eq:DiffEquationCyclotron}
\end{equation}
where 
\begin{equation}
	\vec{\Omega}_c = -\frac{e}{m\gamma}\left(\vec{B}-\frac{\gamma^2}{\gamma^2-1}\frac{\vec{\beta}\times\vec{E}}{c}\right)
\label{eq:CycltronFrequency}
\end{equation}
is the cyclotron frequency.
The relative spin precession $\vec{\Omega}$ of a muon in a storage ring with an electric field $\vec{E}$ and magnetic field $\vec{B}$ is then given by:

\begin{equation}
	\vec{\Omega}=\vec{\Omega}_0-\vec{\Omega}_c=\frac{q}{m}\left[a\vec{B}-\frac{a\gamma}{\left(\gamma+1\right)}\left(\vec{\beta}\cdot\vec{B}\right)\vec{\beta}-\left(a+\frac{1}{1-\gamma^2}\right)
	\frac{\vec{\beta}\times\vec{E}}{c}\right].
\label{eq:omegaMu1}
\end{equation}
which is the known T-BMT equation~\cite{Bargmann1959} when replacing $q=-e$. The presence of the EDM adds a second term

\begin{align}
	\vec{\Omega}=\vec{\Omega}_0-\vec{\Omega}_c=&\frac{q}{m}\left[a\vec{B}-\frac{a\gamma}{\left(\gamma+1\right)}\left(\vec{\beta}\cdot\vec{B}\right)\vec{\beta}-\left(a+\frac{1}{1-\gamma^2}\right)
	\frac{\vec{\beta}\times\vec{E}}{c}\right] \nonumber \\
	%&+\frac{\eta q}{2m}\left[\vec{\beta}\times\vec{B}+\frac{\vec{E}}{c}-\frac{\gamma}{(\gamma+1)c}\left(\vec{\beta}\cdot\vec{E}\right)\vec{\beta}\right].
	&+\frac{\eta q}{2m}\left[\vec{\beta}\times\vec{B}+\frac{\vec{E}}{c}-\frac{\gamma c}{(\gamma+1)}\left(\vec{\beta}\cdot\vec{E}\right)\vec{\beta}\right].
\label{eq:omegaMuWithEDM}
\end{align}

The first line of equation\,\eqref{eq:omegaMuWithEDM}, is the anomalous precession frequency $\omega_{\rm a}$, the difference of the Larmor precession and the cyclotron precession. 
The second line is the precession $\omega_{\rm e}$ due to an EDM coupling to the relativistic electric field of the muon moving in the magnetic field $\vec{B}$, oriented perpendicular to $\vec{B}$. 
In the case that momentum, magnetic field, and electric field form an orthogonal basis, the scalar products of momentum with fields, $\vec{\beta}\cdot\vec{B}=\vec{\beta}\cdot\vec{E}=0$, drop out. A special configuration was chosen for the E821 experiment; muons with a so-called ``magic'' momentum of $p_{\rm magic} = m/\sqrt{a} = \SI{3.09}{GeV/}c$ were used, simplifying equation~\eqref{eq:omegaMuWithEDM} on the reference orbit to

\begin{equation}
	\vec{\Omega}=\frac{q}{m}\left[a\vec{B}+\frac{\eta}{2}\left(\vec{\beta}\times\vec{B}+\frac{\vec{E}}{c}\right)\right],
\label{eq:MagicOmega}
\end{equation}
making the anomalous precession frequency independent of electric fields needed for steering the beam. 
In the presence of a muEDM the precession plane is tilted out of the orbital plane defined by the movement of the muon in this ``magic'' configuration. 
Hence, a vertical precession ($\vec{\omega}_{\rm e} \, \bot \, \vec{B}$) with an amplitude proportional to the EDM with a frequency $\vec{\omega}_{\rm e}$ phase shifted by $\pi/2$ with respect to the horizontal anomalous precession would becomes observable. 
Another effect of an EDM is the increase of the observed precession frequency 

\begin{equation}
\Omega=\sqrt{\omega_{\rm a}^2 + \omega_{\rm e}^2}.
\label{eq:SumOfPrecession}
\end{equation}

\subsection{The frozen-spin technique}
\label{sec:frozenspin}
The experimental setup proposed for this dedicated search for an EDM of the muon is based on ideas and concepts discussed in \cite{Farley2004PRL,Adelmann2010JPG}. \newline 
The salient feature of the proposed search for this hypothetical muon EDM is the exploitation of the large electric field $\vec{E}^\ast = \gamma c\vec{\beta}\times \vec{B}\approx \SI{1}{GV/m}$ in the rest frame of the muon, while canceling the effect of the anomalous moment by a meticulously-chosen electric field. Here, as in the remainder of the document, fields in the rest frame of the particle will be indicated by an $\ast$ while all other notation indicate fields in a laboratory frame.
The anomalous precession term in equation~\eqref{eq:omegaMuWithEDM} can be set to zero by applying an electric field such that

\begin{equation}
		a\vec{B} = \left(a-\frac{1}{\gamma^2-1}\right)\frac{\vec{\beta}\times\vec{E}}{c}.
\label{eq:FrozenSpinCondition}
\end{equation}
In the idealized case of $\vec{\beta}\cdot\vec{B}=\vec{\beta}\cdot\vec{E}=0$, and $\vec{B}\cdot\vec{E}=0$ we find that $E_{\rm f}\approx aBc\beta\gamma^2$. By selecting the exact field condition of equation~\eqref{eq:FrozenSpinCondition}, the cyclotron precession frequency is modified such that the relative angle between momentum vector and spin remains unchanged if $\eta =0$, hence it is ``frozen''. In the presence of an electric dipole moment the change in polarization is described by
\begin{equation}
	\Diff{\vec{\Pi}}{t} = \vec{\omega}_e\times\vec{\Pi},
\label{eq:DiffEqPolEDM}
\end{equation}
where 
\begin{align}
	\vec{\omega}_e &= \frac{\eta q}{2m}\left[\vec{\beta}\times\vec{B}+\frac{\vec{E}_{\rm f}}{c}\right] \nonumber \\
								&= \frac{2d_\mu}{\hbar}\left(\vec{\beta}c\times\vec{B}+\vec{E}_{\rm f}\right)
\label{eq:EDMFrequency}
\end{align}
is the precession frequency due to the electric dipole moment of the muon. For the idealized case, see above, this results in a vertical build-up of the polarization

\begin{align}
	|\vec{\Pi}(t)| = P(t) &= P_0\sin\left(\omega_e t\right) \\
	& \approx P_0\omega_e t \nonumber \\
	& \approx 2 P_0 \frac{d_\mu}{\hbar}\frac{E_{\rm f}}{a\gamma^2}  t. 
	\label{eq:polarizationEDM}
\end{align}
From the slope 
\begin{equation}
	\Diff{P}{d_\mu} = \frac{2P_0E_{\rm f} t}{a\hbar\gamma^2}
\label{eq:EDMSlope}
\end{equation}
multiplied by the mean analysis power of the final polarization, $A$, we calculate the sensitivity as
\begin{equation}
		\sigma(d_\mu)=\frac{a\hbar\gamma}{2P_0E_{\rm f}\sqrt{N} \tau_\mu A},
\label{eq:EDMsensitivity}
\end{equation}
for a search of the muon EDM by replacing $t$ with the mean free laboratory lifetime of the muon in the detector
 $\gamma\tau_\mu$ and scaling by $1/\sqrt{N}$ for the Poisson statistics of $N$ observed muons. The initial polarization $P_0>0.93$ of a beam of muons from backward decaying free pions was measured for a momentum of $\SI{125}{MeV/}c$ muons on $\mu$E1 beam line, see Sec.\,\ref{sec:muE1}. 
For the mean decay asymmetry we take $A=0.3$. 
Hence the EDM sensitivity for a single muon is $\sigma(d_\mu)\approx \SI{1e-16}{\ecm}$, assuming a magnetic field of $B=\SI{1.5}{\tesla}$, which in turn results in an electric field for the  frozen spin condition of $E_{\rm f}=\SI{0.96}{MV/m}$. 
At $\mu$E1 beam line a total muon flux of $\SI{2e8}{\mu^{+}/\second}$ was reported before~\cite{Adelmann2010JPG}. 
Injection simulations indicate a 0.14\% efficiency for lateral injection without material. 
With thin aluminum electrodes this reduces further by a factor 14 to \SI{1e-4}, c.f.\ Sec.~\ref{sec:LateralInjection}.
These numbers indicate that one could store one muon at a time at a rate of $1/\left(\gamma\tau +\langle t_\mathrm{d} \rangle \right) = \SI{18}{kHz}$, where $\langle t_\mathrm{d} \rangle=\SI{50}{\micro \second}$ is the mean waiting time between two successive measurements. Assuming 200 days per year for data taking, this results in a total of $3.2\times10^{11}$ detected positrons per year which in turn yields a sensitivity of $\sigma(d_\mu)\approx \SI{2e-22}{\ecm}$.

We also investigate the option of a vertical injection as described in \cite{Iinuma2016NIMA}, see Sec.~\ref{sec:3DInjection}. The clear advantage is that the muons do not have to pass several times through electrodes, as the lateral injection. Further, the deployment of a magnetic field of up to \SI{3}{T} seems better feasible, as it results in a larger electric field of $E_{\rm f}\approx \SI{2}{MV/m}$, which can be deployed more easily in this scheme as the injection channel is moved far away from the electric field. \newline

To avoid a triggered magnetic field kick, we also investigated a scenario deploying the vertical 3D-injections and avoiding the storage of the muons on a stable orbit altogether. Instead we let them drift through the frozen field configuration on a helix. 
In this case, all muons which can be injected also contribute to the final sensitivity, as we do not have to wait for the decay of each muon before the next is admitted to the experiment. 
However, the larger the drift angle, this means the velocity along the solenoid axis, the shorter the mean time required to pass through the frozen-field region. As a consequence, the fraction of decays within the frozen field region is reduced, although the rate of injected muons increases to $\SI{234}{kHz}$, resulting in an effective mean storage time of about \SI{40}{ns}. This in turn results in an annual sensitivity of $\sigma(d_\mu)\approx \SI{3e-21}{\ecm}$ in the case of a \SI{3}{T} magnetic field.

The most sensitive scenario would be to apply a vertical magnetic kick to the injected muon to store it on a stable orbit as in the classical storage-ring concept. 
On the one hand, losses due to multiple scattering on the electrodes and injection channel vanish compared to the lateral injection scheme. While on the other hand, the requirement of triggering the injection procedure relaxes to about \SI{50}{ns}. Combined with a magnetic field of \SI{3}{T} this results in a sensitivity of

\begin{equation}
	\sigma(d_\mu)\approx \SI{6e-23}{\ecm},
\label{eq:EDMsensitivityNumeric}
\end{equation}
as detailed in Sec.~\ref{sec:3DInjection}.

\subsection{Compact storage ring with lateral injection}
\label{sec:storagering}

A straw-man idea for the lateral injection approach of the experiment is shown in Figure~\ref{fig:StorageRingMuEDM}. A muon beam from PSI's $\mu$E1 (or $\pi$E1) beam line is first collimated to limit the vertical divergence of the beam. 
It is then guided to the central region of a weak-focusing magnet through a magnetic channel (injection channel). 
Upon entering the magnet, the muon is displaced from its storage orbit by a few centimeters. 
Without any magnetic/electric steering, it will come back to the same place, hit the magnetic channel and scatter out of the storage ring.

As the period of the muon cyclotron motion is around 10\,ns, conventional beam steering techniques, for example the one-turn-pulsed-magnetic kicker utilized by the Muon $g-2$ experiment at Fermilab~\cite{Grange:2015fou}, are not directly applicable here. The resonance injection technique~\cite{Takayama:1987} that relies on exciting half-integer resonances in beam betatron motion in the storage ring was demonstrated in compact electron storage rings and can be applied here.

After a quarter of a turn, the beam moves into the inner radius region and overlaps with the field region of a perturbator (PB)~\cite{Takahashi:1987sp}. 
The PB is ramped up before the beam arrives using an accelerator trigger, so that the effective field index of the storage ring is close to $n = 0.72$ ($Q_x = \sqrt{1-n}=0.5$, condition for half-integer resonance).
It is then damped over 20 turns ($\approx200$\,ns) until the muon is relaxing onto its storage orbit. The effective magnetic field index of the system is reduced from 0.72 to 0.25 over the same time period. 
Once the muon is stored, the required radial electric field $E_{r}\sim aBc\beta\gamma^2$ that can be applied by means of two concentric cylinder electrodes will ``freeze'' the muon spin relative to the
momentum as the muon circulates in the storage ring.

\begin{figure}
\centering
\subfloat[][]{
\includegraphics[width=0.45\textwidth]{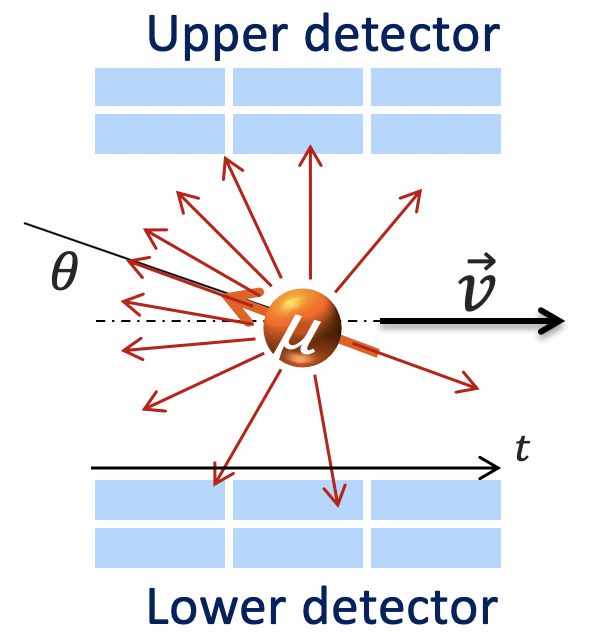}\label{fig:UpDownAsymPic} }
\hfill
\subfloat[][]{
\includegraphics[width=0.5\textwidth]{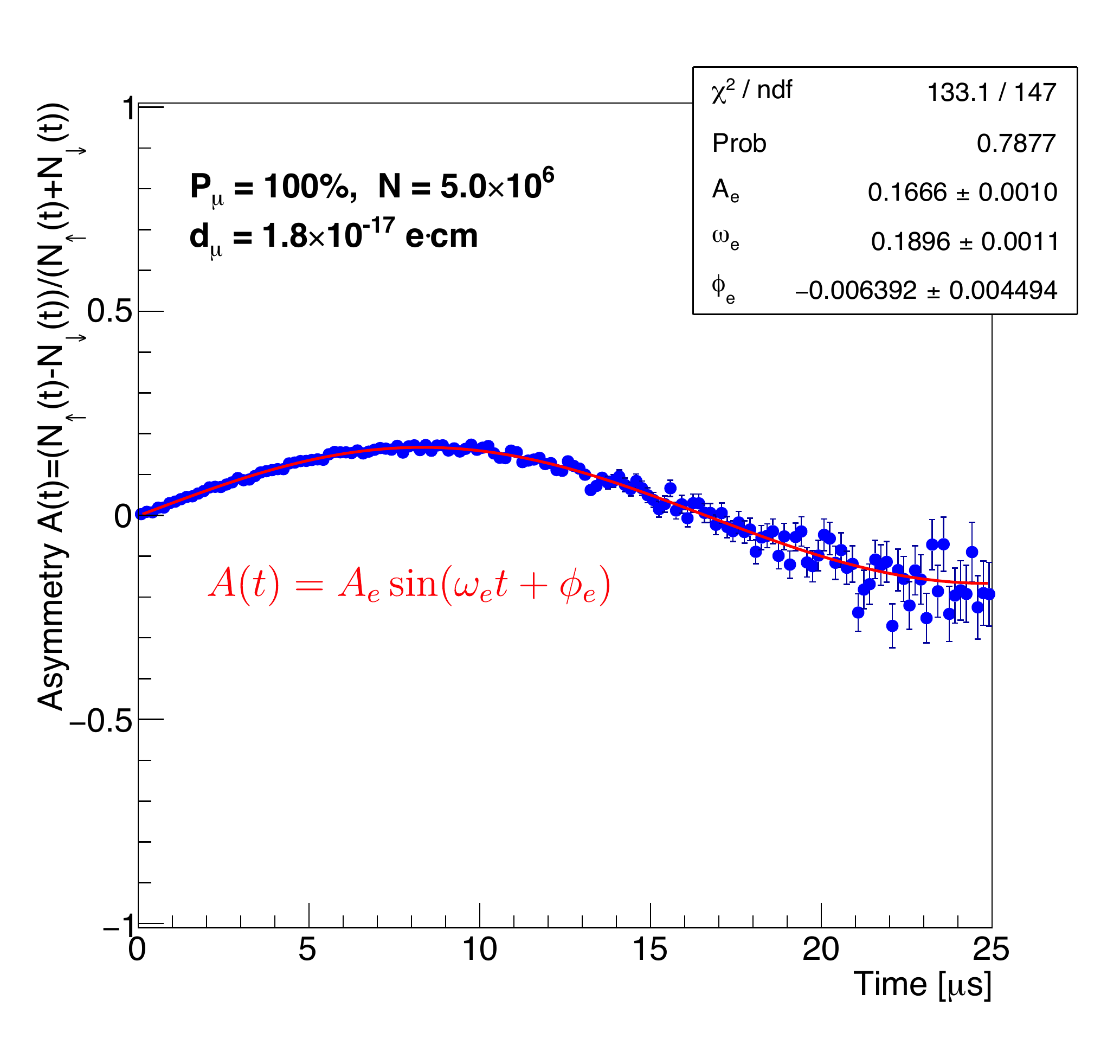}\label{fig:UpDownAsymGraph} }
\caption{(a)~A detection concept for the muon EDM. (b)~Simulated ideal up-down asymmetry plot assuming a large muon EDM of $d_\mu=\SI{1.8e-17}{\ecm}$. The red line is the fit to simulated data points.}\label{fig:UpDownAsym}
\end{figure}
\begin{figure}
	\centering
	\subfloat[][]{
		\includegraphics[width=0.45\textwidth]{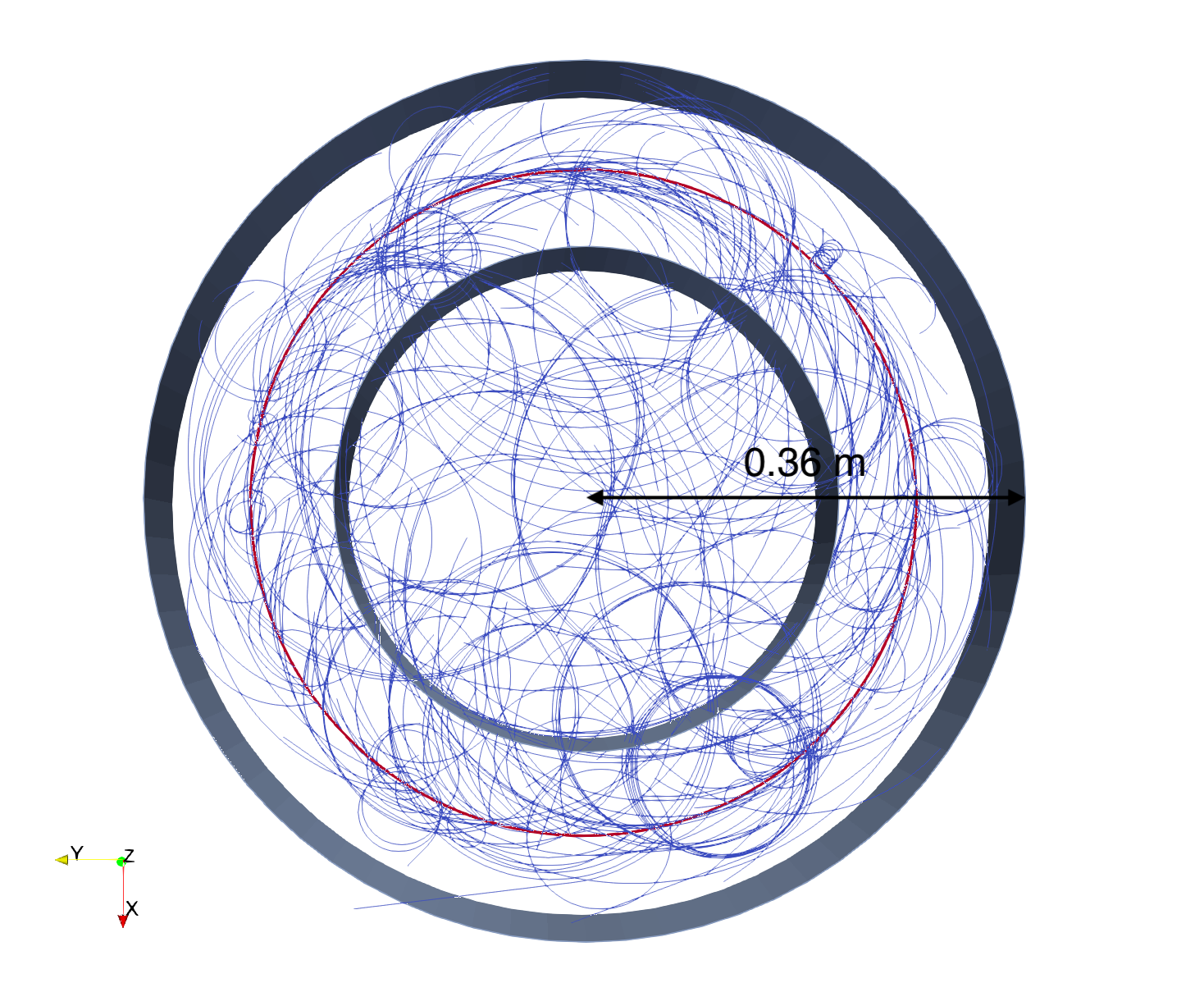}\label{fig:StorageRing_SimTop} }
	\hfill
	\subfloat[][]{
		\includegraphics[width=0.45\textwidth]{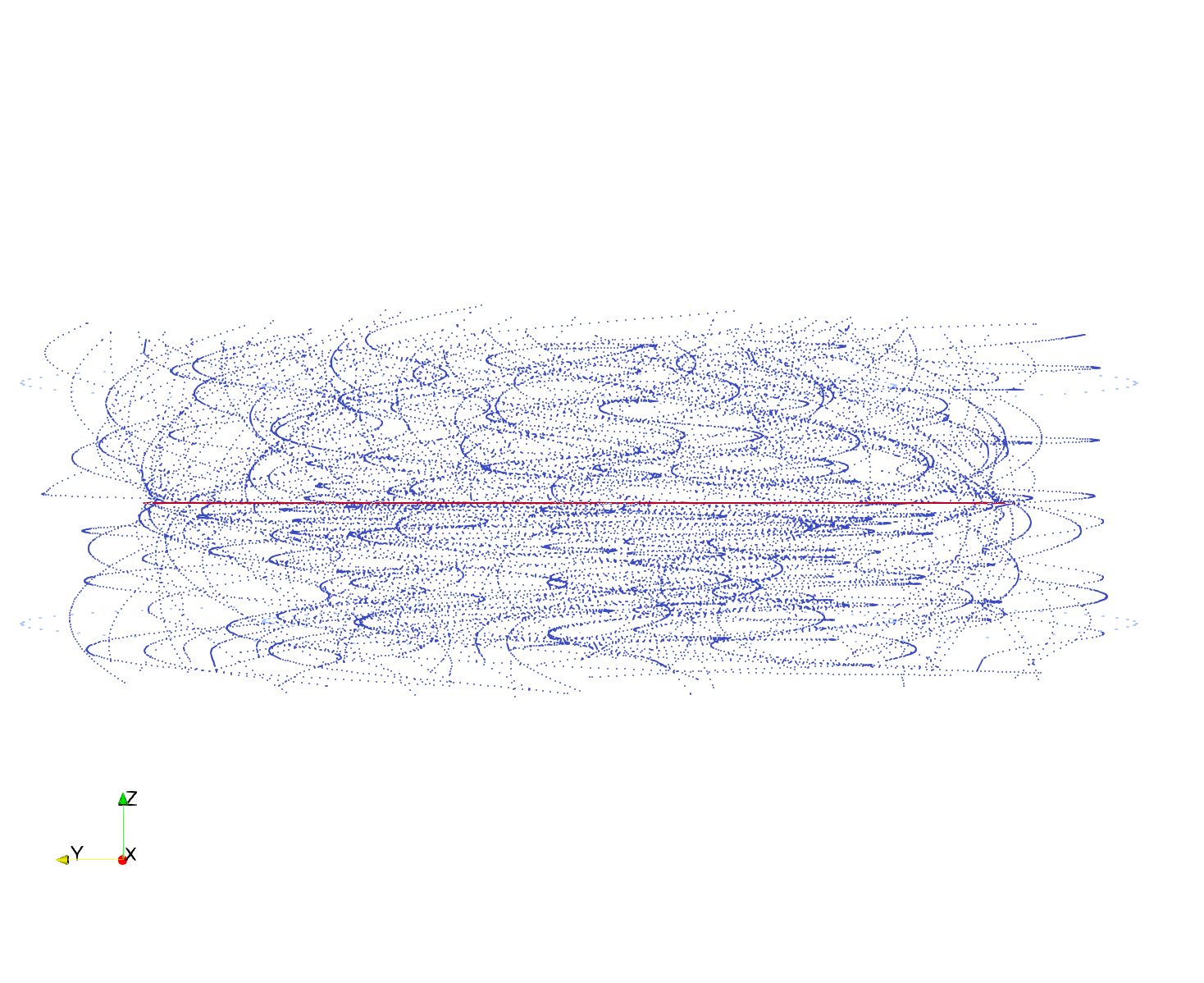}\label{fig:StorageRing_SimSide} }
	\caption{(a)~Top view and (b)~side view of the ideal simulation with $200$ muons at $125$~MeV/$c$. The stored muon orbits are shown in red and the decay positron trajectories are in blue. The dark rings, changing their width are a perspective view of the electrodes.}\label{fig:StorageRing_Sim}
\end{figure}
For the detection of decay positrons, an EDM detector can be installed at the top and bottom of the muon orbital plane. As the muon beam is circulating in the storage ring, the spin will follow the direction of the muon momentum if the muon has no EDM\@. 
If an EDM exists for the muon, the spin will slowly precess out of the muon orbital plane. 
Thus an observable up-down asymmetry that oscillates with time, with a frequency directly proportional to the muon EDM, can be observed. 
In the case of a muon EDM smaller than the current limit, a slow increase in the up-down asymmetry in Figure~\ref{fig:UpDownAsymGraph} is expected, as the amount of spin precession out of the orbital plane is limited by the muon lifetime. 
As most of the positrons will curl into the center of the storage ring as shown in Figure~\ref{fig:StorageRing_Sim}, a positron tracker made of scintillating fibers and depleted monolithic active pixel sensors (DMAPS) can be installed in the inner part of the storage ring to track the positrons. 
It can be used to measure the residual anomalous precession signal and fine-tuning the radial E-field to reach the ``frozen-spin" condition, and to discriminate up and downward tracks for the EDM analysis.
\newpage
\subsubsection{Simulation of lateral injection into a compact storage ring}
\label{sec:LateralInjection}
\begin{figure}%
\centering
\subfloat[][]{\includegraphics[width=0.46\columnwidth]{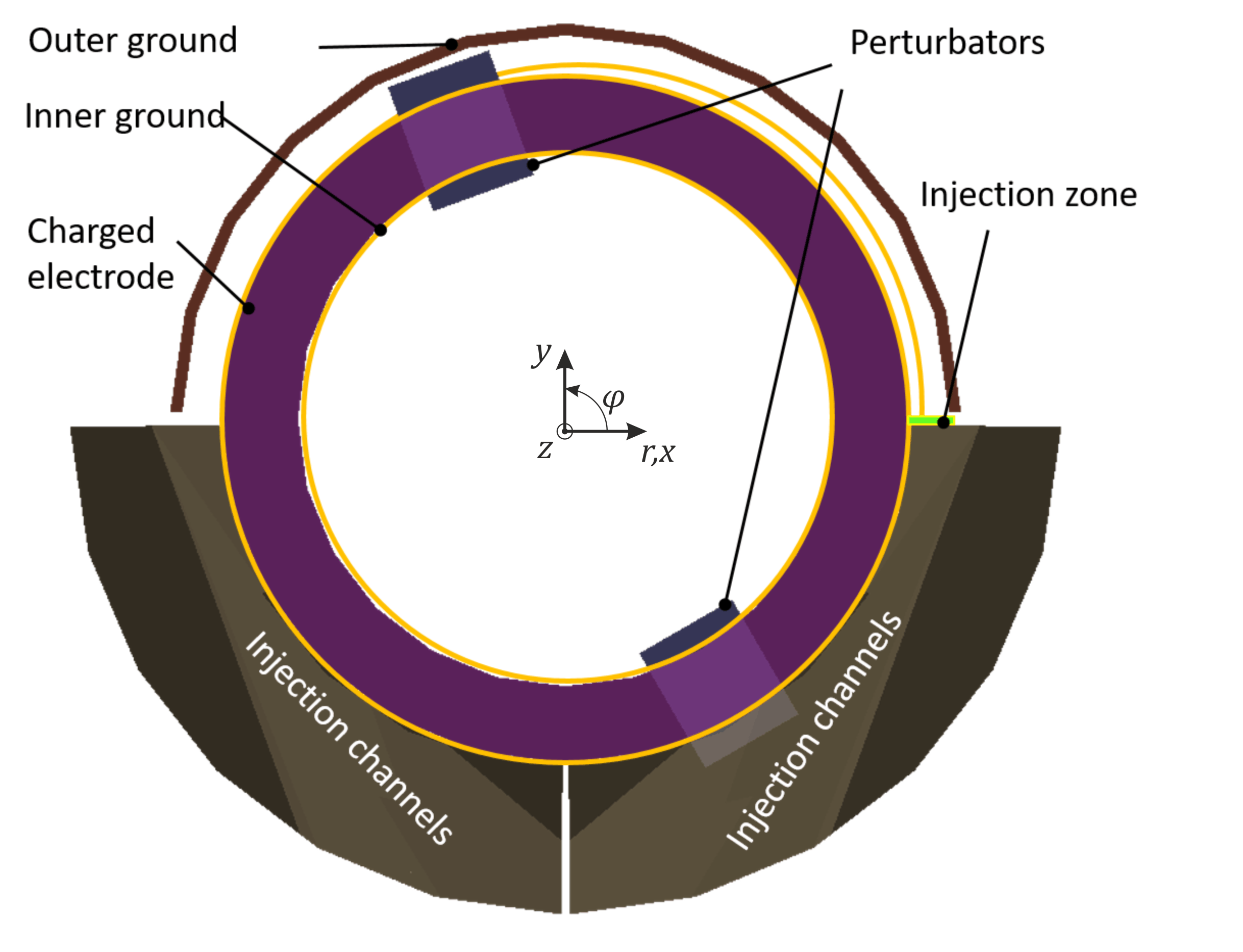}\label{fig:SimGeometry}}%
\hfill
\subfloat[][]{\includegraphics[width=0.46\columnwidth]{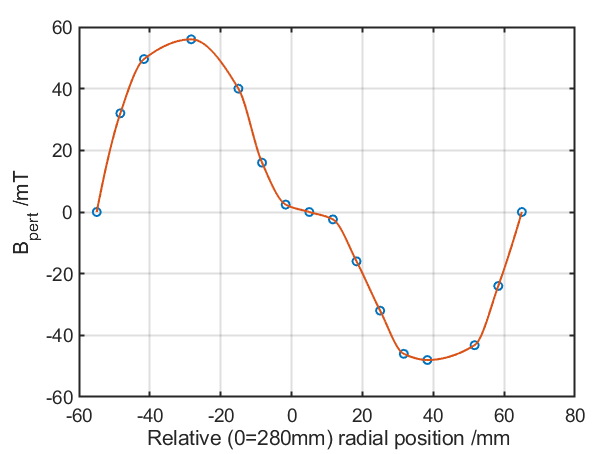}\label{fig:Pertfield}}%
\caption{Geometry of the storage ring~(a) used for the simulation study of lateral injection. The origin of the coordinate system is the center of the reference trajectory in the $z=0$ plane with radius $r=\SI{280}{mm}$. In the simulation the muons are created in the injection zone which extends for \SI{20}{mm} radially and \SI{10}{mm} vertically just outside of the negative charged high-voltage electrode. A second injection channel is required for counterclockwise measurements. Along the nominal orbit two perturbation fields~(b) are applied during injection and ramped to zero within \SI{150}{ns}. Note that for counterclockwise injection a second pair of perturbators is required, but not shown here.}%
\label{fig:SimLateralInjection1}%
\end{figure}

The lateral injection of muons into a weakly-focusing magnetic field of \SI{1.5}{T} was simulated using \textsc{G4beamline}~\cite{Roberts2011,Roberts2012}. Figure~\ref{fig:SimGeometry} shows the implemented geometry, the weakly-focusing magnetic field was modeled, using the formalism defined in\,\cite{Abel2019PRA}, as 

%\begin{equation}
%	\vec{B}(\rho,z)= 
%		\begin{pmatrix}
%		0 \\
%		0 \\
%		B_{0z}
%	\end{pmatrix} + G_{20}
%	\begin{pmatrix}
%		0 \\
%		-\rho z\\
%		z^2-\rho^2/2
%	\end{pmatrix},
%\label{eq:}
%\end{equation}
\begin{equation}
	\vec{B}(r,z)= 
		\begin{pmatrix}
		0 \\
		0 \\
		B_{0z}
	\end{pmatrix} + G_{20}
	\begin{pmatrix}
		-r z\\
		0 \\
		z^2-r^2/2
	\end{pmatrix},
\label{eq:}
\end{equation}
where $B_{0z}=-\SI{1.509}{T}$ and $G_{20}=-0.018/280^2$\,${\rm T/mm^2}$, which results in $B_{0z}=\SI{-1.5}{T}$ on the reference orbit with $r=\SI{280}{mm}$ and $z=0$. The radial electric field of $E_{\rm f} = \SI{0.962}{MV/m}$ is applied within a cylindrical volume of $\SI{243}{mm}<r<\SI{313}{mm}$ and $\SI{-59.5}{mm}<z<\SI{59.5}{mm}$ defined by two thin aluminum foils of $d=\SI{20}{\micro\meter}$ thickness. 
The muons for the injection simulations are created at $\varphi=0$ (see Figure~\ref{fig:SimGeometry}), $\SI{313}{mm}<r<\SI{333}{mm}$ and $\SI{-10}{mm}<z<\SI{10}{mm}$, with a divergence of $\SI{-10}{mrad}<r'<\SI{10}{mrad}$ and $\SI{-10}{mrad}<z'<\SI{10}{mrad}$. The perturbation fields, shown in Figure~\ref{fig:Pertfield}, expand over $\Delta\phi=\pm\SI{10}{\degree}$ at $\phi=\SI{110}{\degree}$ and $\phi=\SI{200}{\degree}$ and are ramped down with a delay of \SI{4}{ns} after creation within \SI{150}{ns}. 
The field shape is a copy of the field published in \cite{Takayama:1987}. In addition to the electrodes generating the frozen-field region, an additional ground electrode will be needed outside of the charged electrode. For the simulation, this was modeled as half cylinder made of copper in the range $\phi=$\SIrange{0}{180}{\degree} at $r=\SI{348}{mm}$ and $\SI{-59.5}{mm}<z<\SI{59.5}{mm}$, while in the sectors $\phi=$\SIrange{180}{360}{\degree} two injection channels made of magnet iron are positioned. 

\begin{figure}%
\centering
\subfloat[][]{\includegraphics[width=0.52\columnwidth]{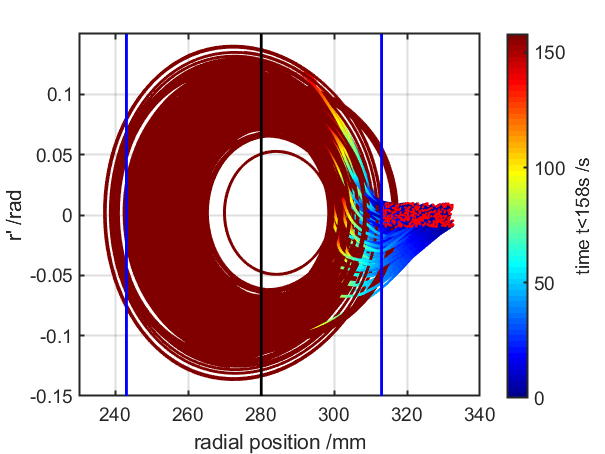}\label{fig:PerfectInjection}}%
\hfill
\subfloat[][]{\includegraphics[width=0.44\columnwidth]{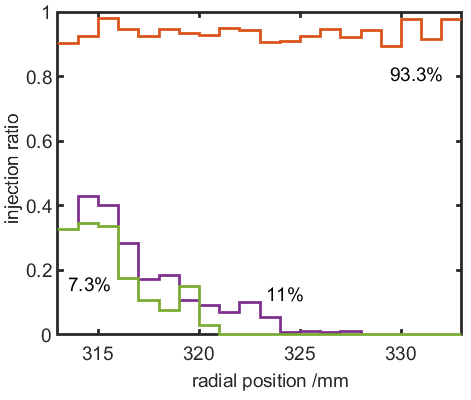}\label{fig:InjectionRatio}}%
\caption{Radial phase-space evolution during injection~(a) for the case that all materials are set to vacuum in the simulation. The track color changes with time from dark blue at $t=0$ to brown $t=\SI{158}{ns}$.  Note, the nominal orbit $r=\SI{280}{mm}$ for muons with $p=\SI{125}{MeV/}c$ is indicated by the black vertical line. The two blue vertical lines indicate the position of the ground and high voltage electrode. Red points depict the initial creation of the simulated muon. The injection ratio~(b) illustrates the effect once materials are included into the simulation. Without material 93\% of the muons decay to a positron and two neutrinos within a volume defined by the electrodes and $-\SI{55}{mm}\leq z\leq \SI{55}{mm}$. The dramatic loss of muons, only 11\% decay within the frozen-spin volume, can be traced back to multiple scattering within the thin aluminum electrode (thickness $d=\SI{20}{\micro\meter}$) during many passages through matter. A reduction of these losses is possible by using even thinner electrodes, however, the addition of a required low-field region from the injection channel in the injection zone reduces the injection efficiency further. In the illustrated case this result in a total injection efficiency of just above 7\%.}%
\label{fig:SimLateralInjection2}%
\end{figure}

The choice for the reference injection phase space was driven by a series of simulations varying the lateral and vertical phase-space parameters. 
In the case when all materials are set to vacuum, losses only occur due to a too large vertical divergence, which could be counteracted by an even stronger weakly-focusing component $G_{20}$. 
Figure~\ref{fig:PerfectInjection} shows the radial phase space evolution using the injection phase space above-mentioned and a kick field which is linearly ramped down within \SI{150}{ns}. 
Without vertical divergence, it was possible to inject nearly 99\% of all muons from a lateral phase space of $28\times17~\mathrm{mm}\!\cdot\!\mathrm{mrad}$ as was also demonstrated in \cite{Adelmann2010JPG}. 
However, as simulations quickly showed, most losses in a realistic configuration occur due to multiple scattering in the many passages of the muon through the high voltage electrode or by hitting the entrance channel. 
Hence, a further refinement of the vertical divergence and adaptation of $G_{20}$ seemed superfluous. 
Figure~\ref{fig:InjectionRatio} illustrates, nearly 90\% of all muons are lost during injection due to multiple scattering once material properties are turned on in the simulation. A change of the electrode design could most probably reduce these losses by using electrodes made of low-$Z$ material, e.g., a thin Kapton foil coated with an even thinner layer of aluminum. However, losses also occur due to a return of the muons into the injection region. 
In the case of the simulation presented here, this leads to another 30\% loss due to passages through the low field area from the injection channel at every turn. 
In total, we observed a loss of 93\% of all created muons in the injection zone, which reduces the injection efficiency by a factor 13.6\@.

\subsection{Stored or continuous measurement using a vertical helix injection}
\label{sec:3DInjection}
\begin{figure}%
\centering
\subfloat[][]{\includegraphics[width=0.49\columnwidth]{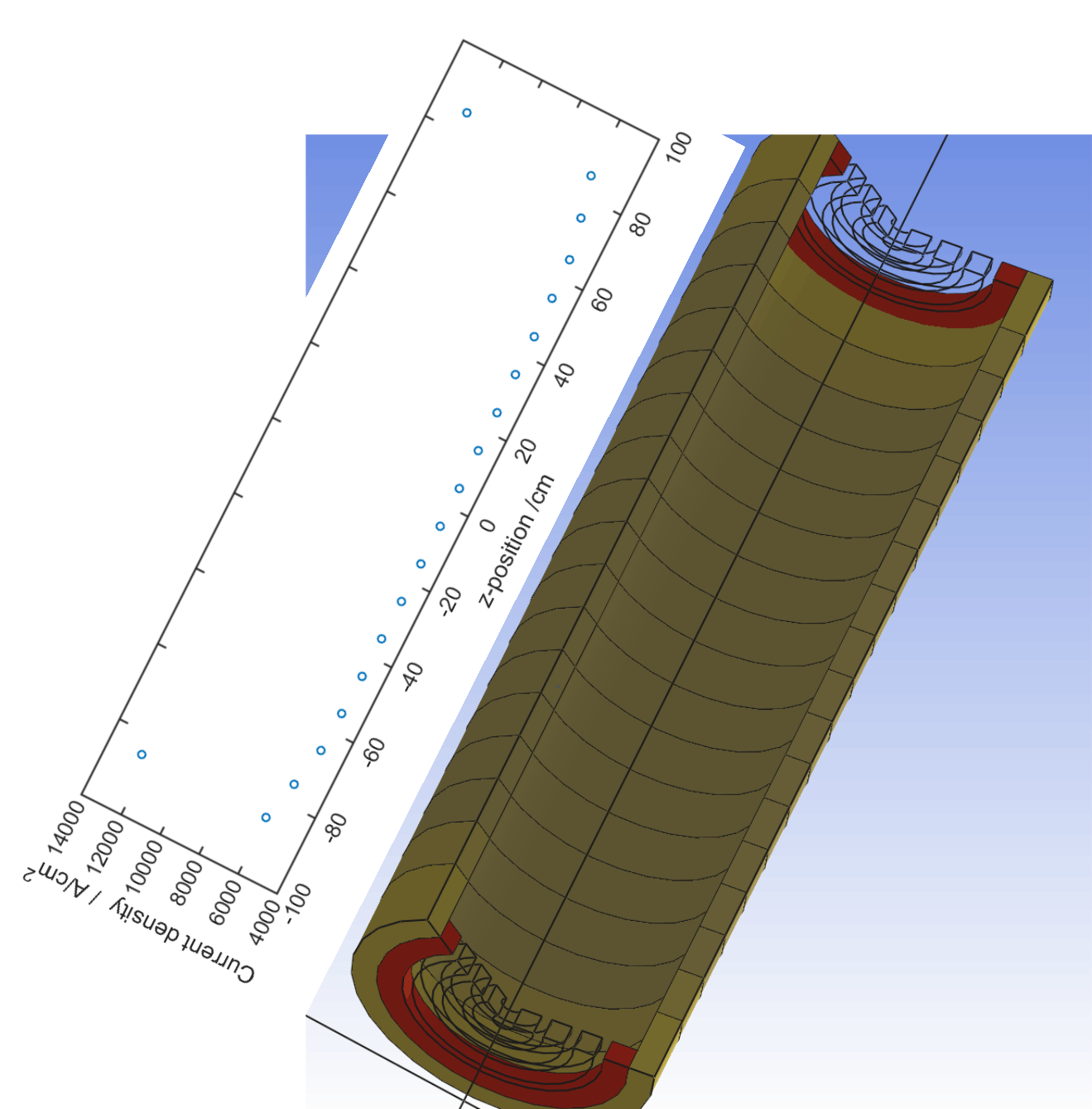}
\label{fig:SolenoidFEM}}
\hfill%
\subfloat[][]{\includegraphics[width=0.47\columnwidth]{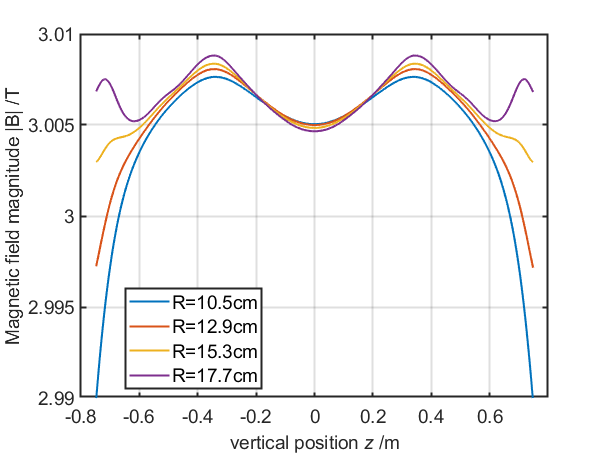}
\label{fig:SolenoidField}}
\caption{(a)~Computer-rendered image of the coil geometry deployed to calculate the magnetic field. The light yellow-grey, and red cylinders are individual coils, the  coil package, used in finite element simulation to create a highly-uniform magnetic field shown in (b). The currents shown in the the inlay were deduced by manual adjustments, with the goal to produce a field that is less than 1\% smaller in the injection area $760>|z|>\SI{740}{mm}$ than in the center $z=0$, and to create a weak-focusing field at $z=0$. (b)~Magnitude of the magnetic field for several radii, $R$; the weak-focusing field is clearly visible in the range $z=$\SIrange{-0.3}{0.3}{m}. Note that the nominal radius of a stored muon is approximately \SI{14}{cm}.}%
\label{fig:UniformSolenoid}%
\end{figure}

An alternative injection, originally proposed and pioneered for the Japanese ($g-2$) project at J-PARC~\cite{Iinuma2016NIMA}, is the injection of muons outside the central and highly uniform magnetic field under a vertical angle $\zeta = \vec{p}_{\parallel}/\vec{p}_\bot$ into a field produced by a solenoid-like coil package. Here $\parallel$ indicates the momentum component parallel to the magnetic field $\vec{B}(r,z)$ on the symmetry axis $r=0~\forall~z$. 
Figure~\ref{fig:SolenoidFEM} shows a possible coil package producing the field shown in Figure~\ref{fig:SolenoidField}. 
This method circumvents the large losses due to multiple passages through material. In combination with a trigger/tagger system upstream, see Sec.\,\ref{sec:tagger} it lends itself well for single-muon storage measurement by applying a vertical magnetic kick as described in~\cite{Iinuma2016NIMA}. The entrance trigger will also set a veto in order to inhibit a second magnetic field pulse during the storage period of the muon. Muons which still enter into the solenoid, will quickly pass through the central region and are stopped far away from the positron detection system. The veto is removed by the detection of a decay positron by rapid scintillating tiles next to the positron tracker, or latest after four laboratory life times of about \SI{14}{\micro s}. The spectrometer is again ready to accept the next muon for storage.

A second option is to operate this configuration continuously without magnetic kick and let the muons drift through the entire field.
%, similar as the trace in \autoref{fig:FlatTopSolenoidField} indicates.
In this case an event-by-event reconstruction will be implemented using a muon tagger, see Sec.~\ref{sec:tagger}, at the entrance providing the injection angle $\zeta$ and start time $t_0$ for each muon. 
In combination with the information of the central positron tracker, the decay vertex and the vertical decay asymmetry can be reconstructed. 

\subsubsection{Stored muons from vertical injection} 
As in the lateral injection case, the vertical injection needs a triggered magnetic field to kick the muons onto a stable orbit in the magnetic-field plane at $z=0$.
%This feature was successfully demonstrated using positrons~\cite{Iinuma2016NIMA}. 
Also, in this case, a fast trigger/tagger system is required to start the magnetic kick. 
However, as demonstrated below in the simulation, Sec.~\ref{sec:3DSimulation},  the kicker needs to be triggered only after \SI{50}{ns}, which is considerably longer than in the lateral case. 
One could use a combination of machine frequency and anti-coincidence between an entrance and veto scintillators inside the injection channel to produce the trigger for the magnetic kick power supply.
As multiple transitions through material could be avoided, we expect a significant gain in injection efficiency once the magnetic weakly-focusing field and the magnetic kick are optimized. First, simulations for the vertical injection were performed and are described in Sec.~\ref{sec:3DSimulation}. More details will be studied soon to optimize the design.  
For now, it seems to be sensible to start with an injection phase space  of about the same as in the lateral injection case: $20 \times 20~\mathrm{mm} \!\cdot\!\mathrm{mrad}$ horizontal and $20 \times 20~\mathrm{mm}\!\cdot\!\mathrm{mrad}$ vertical (both FWHM). 
Together with the measurements presented in \autoref{sec:muE1} and the preliminary injection efficiency deduced in simulations, this results in an injection efficiency of \SI{0.5e-3} and a positron detection rate of about \SI{60}{kHz}. Which in turn translates into the statistical sensitivity given in equation~\eqref{eq:EDMsensitivityNumeric}, as the losses due to multiple scattering can be eliminated, and the dominant factor remains the mean storage time of about \SI{3.4}{\micro s}.

\subsubsection{Simulation of vertical injection}
\label{sec:3DSimulation}
\begin{figure}%
\centering
%\subfloat[][]{
%\includegraphics[width=0.47\columnwidth]{DriftAngleEtaVsGamma}\label{fig:ZetaVsGamma}}
%
%\hfill
%\subfloat[][]{
\includegraphics[width=0.77\columnwidth]{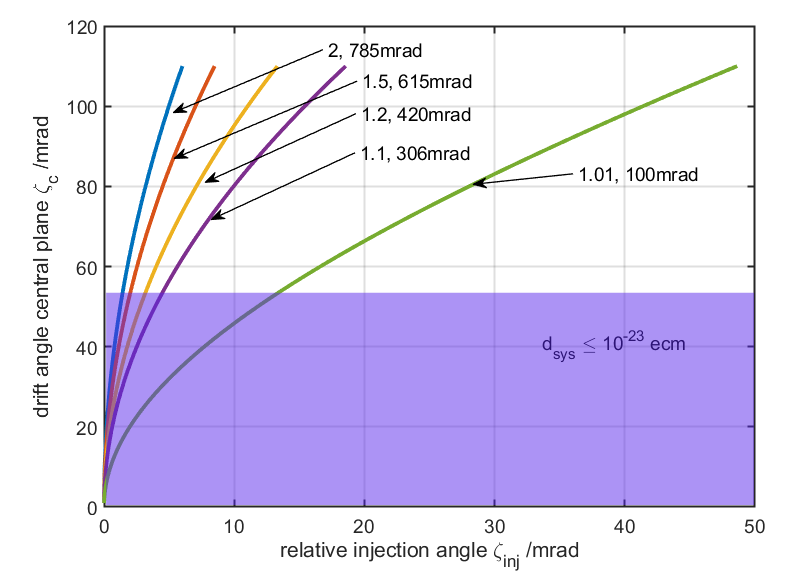}%\label{fig:InjAngleVsDriftAngle}}%
\caption{
%(a)~Maximal allowed drift angle $\zeta$ versus $\gamma$-factor, see equation~\eqref{eq:29}. Note that in the case of a frozen-spin scenario, any field dependents vanishes. (b)~
Plot of injection angle offset $\delta\zeta_{\rm inj}$ versus drift angle $\zeta_{\rm c}$ in the center of the magnet. Each data pair indicates the ratio $B_{\rm c}/B_{\rm inj}$, and the nominal injection angle $\zeta_{\rm inj}$, which results in $\zeta_{\rm c}=0$.}%
\label{fig:InjectionAndDrift}
\end{figure}

The magnetic field and the trajectory of a muon are intimately linked to each other, and essentially it needs many iterations to arrive at an optimized magnetic-field configuration to permit an efficient injection and a stable and well-defined central orbit within the region with frozen-spin condition. The following considerations define the starting point of the initial magnetic field and trajectory simulations presented below.
\begin{enumerate}
	\item Magnetic adiabatic collimation indicates that an initial beam divergence $\zeta_{\rm inj}$ in the injection area (where $B_{\rm inj}$), will be increased to 

\begin{equation}
   \zeta_{\rm c} ={\arccos}\left(\sqrt{\frac{B_{\rm inj}}{B_{\rm c}}\cos(\zeta_{\rm inj})^2}\right),
\label{eq:MAC}
\end{equation}
where $\zeta_{\rm c}$ is the divergence in the central plane with magnetic field $B_{\rm c}$, see also Figure~\ref{fig:InjectionAndDrift}. 
	\item For the storage of the muon in the central plane a weak-focusing field is required with a large vertical acceptance,
	\item and the pulsed magnetic field should efficiently ``stop'' the vertical drift of muons with an as large as possible vertical divergence.
\end{enumerate}

\begin{figure}%
\centering
\subfloat[][]{\includegraphics[width=0.43\columnwidth]{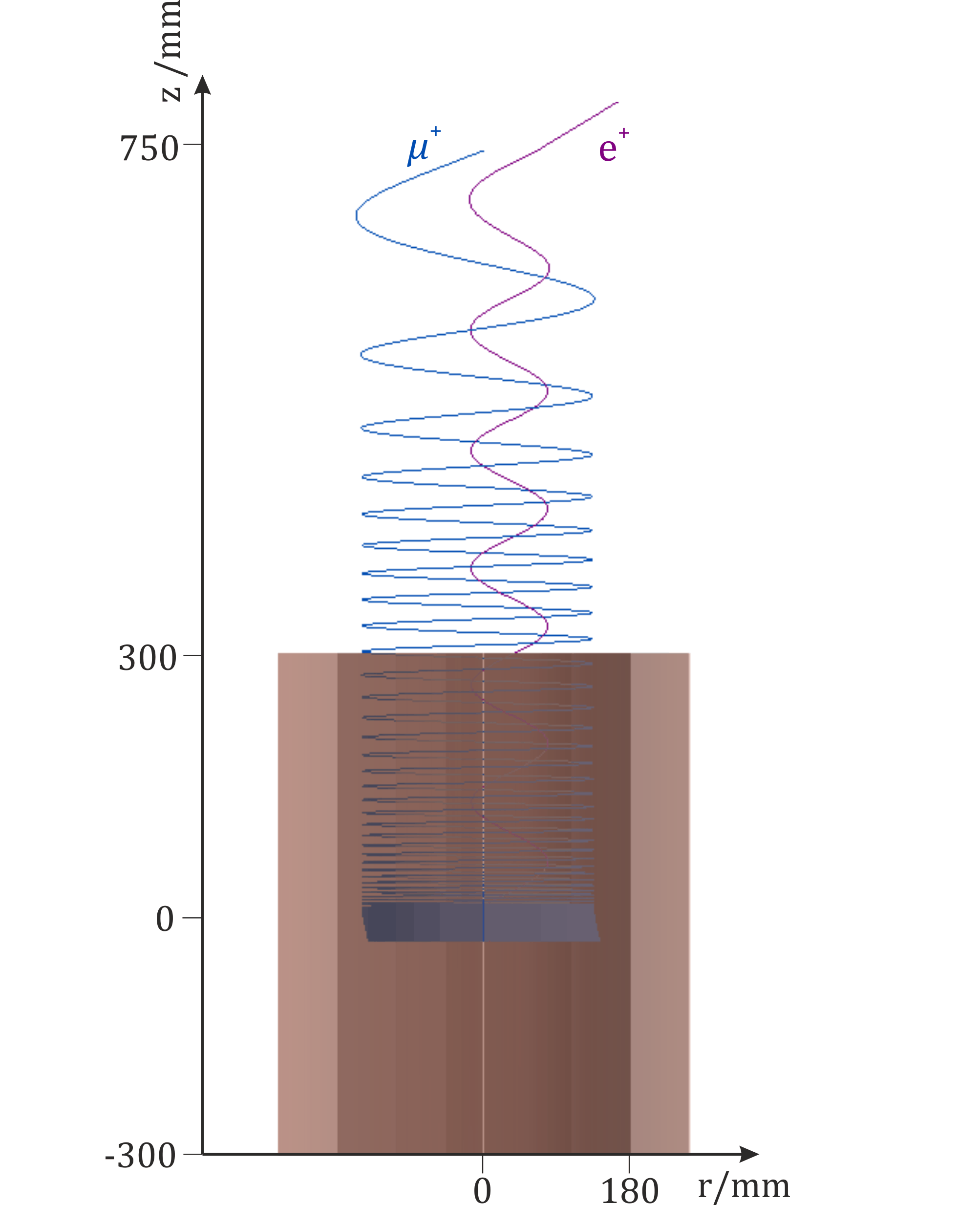}
\label{fig:3DInjectionViewSide}}
\hfill%
\subfloat[][]{\includegraphics[width=0.48\columnwidth]{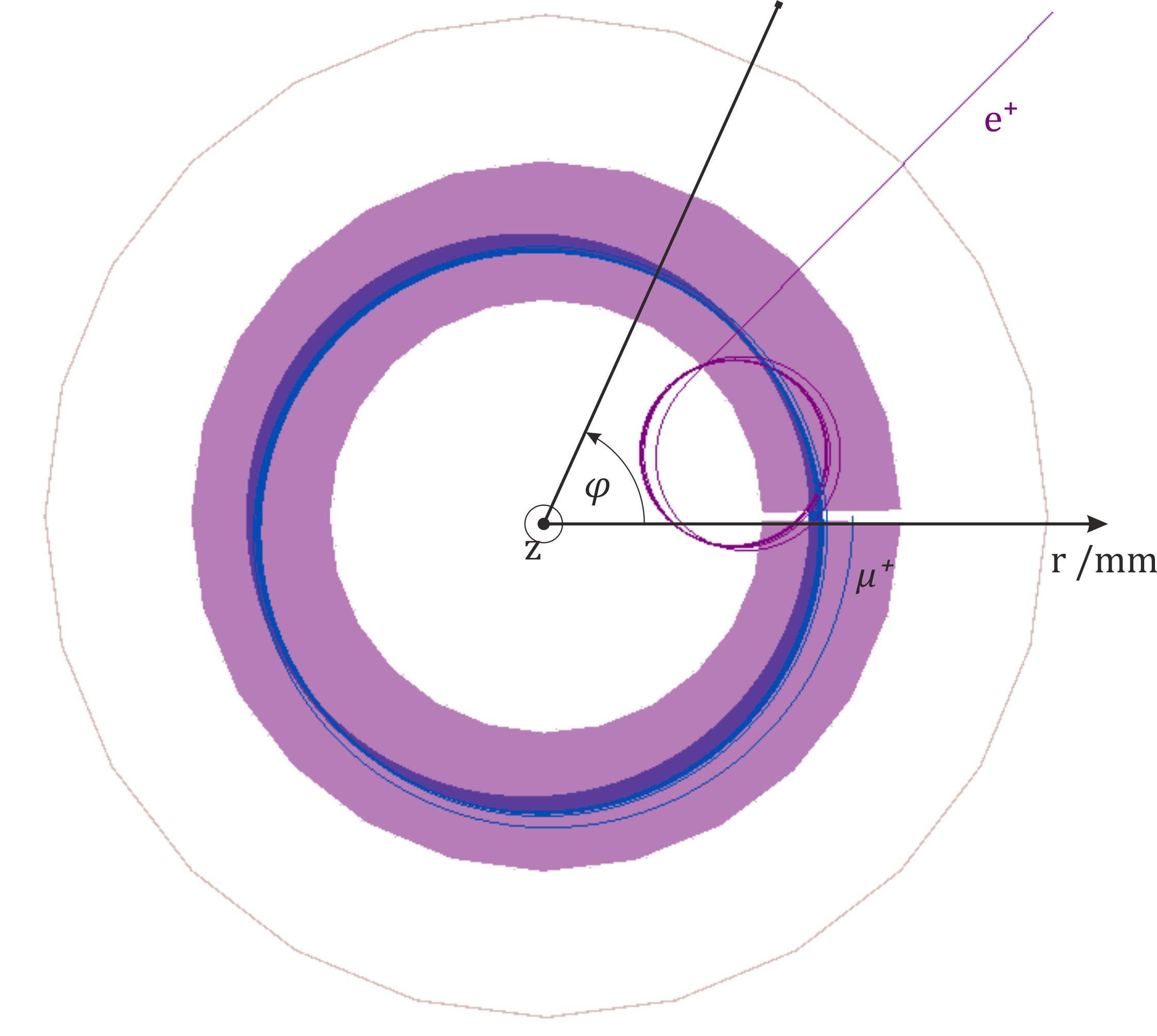}
\label{fig:3DInjectionViewTop}}
\caption{(a)~Simulation images of a single muon injected into the magnetic field shown in Fig.~\ref{fig:SolenoidField}, including electrode system (dark brown cylinders) and electric fields for the frozen-spin configuration. The muon (blue, large radius helix) enters from the top and is kicked when entering the region of the magnetic kicker, $-\SI{300}{mm}<z<\SI{300}{mm}$. A sinusoidal kick stops the vertical drift and the muon is stored in the central region, here $-\SI{20}{mm}<z<\SI{20}{mm}$, until it decays to a positron (dark purple, small radius helix). Note that no detection system is present and the positron escapes the system. (b)~Top view of (a). The electric-field region is defined by the purple zone.}%
\label{fig:InjectionG4Beamline}%
\end{figure}

For the finite element calculation of the uniform solenoid field shown in Figure~\ref{fig:UniformSolenoid} we used \textsc{Agros2D}\,\cite{Karban2013} while for post processing \textsc{Matlab} was used to create magnetic-field maps. These field maps were then used in \textsc{G4beamline} to simulate trajectories of muons and decay positrons.
Figure~\ref{fig:InjectionG4Beamline} shows the side and top view of an injected muon into the central field region including the frozen-spin electric field, $E_{\rm f}\approx \SI{2}{MV/m}$, and the decay positron that leaves the solenoid to the top. 

\begin{figure}%
\centering
\subfloat[][]{\includegraphics[width=0.47\columnwidth]{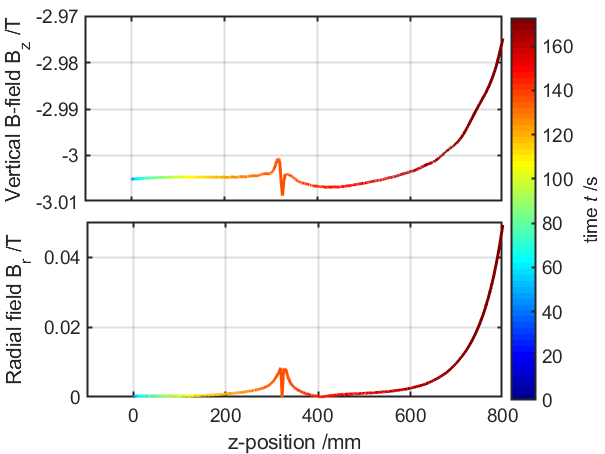}
\label{fig:3DFieldAlongTrajectory}}
\hfill%
\subfloat[][]{\includegraphics[width=0.47\columnwidth]{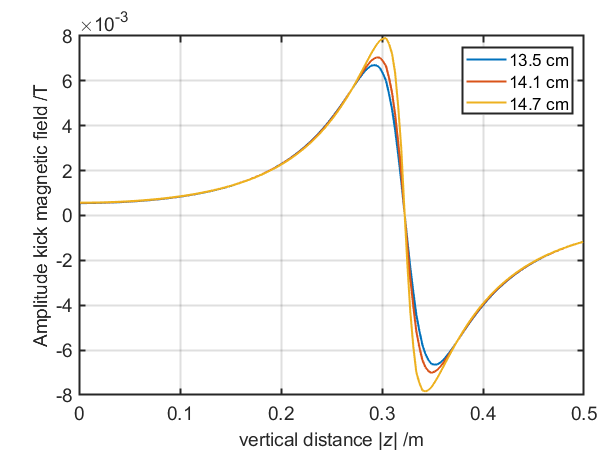}
\label{fig:3DBKickField}}
\caption{(a)~Magnetic field along the reference trajectory, top vertical, bottom radial. The trajectory is from a time-reversed simulated using a positrons. Within the first \SI{48}{ns} the positron remained on a stable orbit at $z=0$. The distinctive feature at $z\approx\SI{320}{mm}$ is the pulsed magnetic kick $\delta t \approx \SI{70}{ns}$ after the pulse was initiated. (b)~Radial-symmetric magnetic-field amplitude $B_{\rm A}$ of the kick used to stop an injected muon for three different radii. The pulse is a half-sine with $B_{\rm kick}(t)=B_{\rm A}\sin(\pi \delta t/T)$, where $T=\SI{100}{ns}$ is the pulse duration. }%
\label{fig:3DFields}%
\end{figure}

In an initial simulation, using the field shown in Figure~\ref{fig:SolenoidField}, we looked at the time-reversed process by generating a positron at $z=0$ and $r=\SI{14}{cm}$ with a momentum of $p=\SI{125}{MeV/}c$. As it had no vertical momentum component, it stayed at $z=0$ until the magnetic kick at $t=\SI{48}{ns}$ was started lasting for $T=\SI{100}{ns}$ with a half sinusoidal period, $B_{\rm kick}(t)=B_{\rm A}\sin(\pi \delta t/T)$. The magnetic field along the positron trajectory is shown in Figure~\ref{fig:3DFieldAlongTrajectory}, while Figure~\ref{fig:3DBKickField} shows the amplitude of the magnetic kick $B_{\rm kick}$. 
The reverse of the vertical ejection angle, $\zeta_{\rm eject} =\SI{89.8}{mrad}$ of the positron at $z=\SI{750}{mm}$, where $B_{rm inj}=B(z=\SI{750}{mm})$ is defined, was used to fix the nominal injection angle for muons incident through the injection channel.  Note that the injection channel was not yet included in the simulation, instead we defined a $20\times20~{\rm mm^2}$ region inclined by the injection angle in which muons with at horizontal and vertical divergence of $\pm\SI{10}{mrad}$ were generated. 

\begin{figure}%
\centering
\subfloat[][]{\includegraphics[width=0.47\columnwidth]{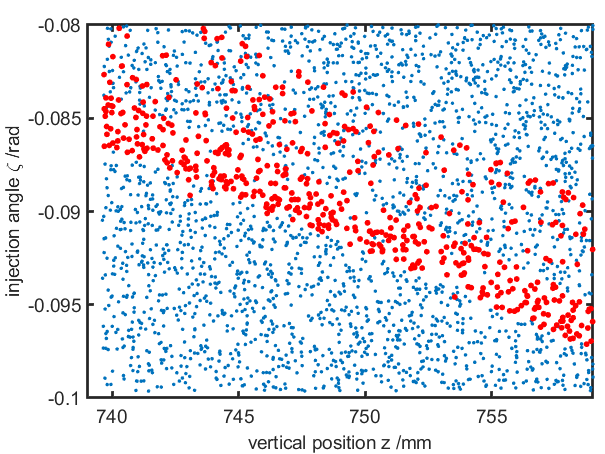}
\label{fig:3DVertAcceptance}}
\hfill%
\subfloat[][]{\includegraphics[width=0.47\columnwidth]{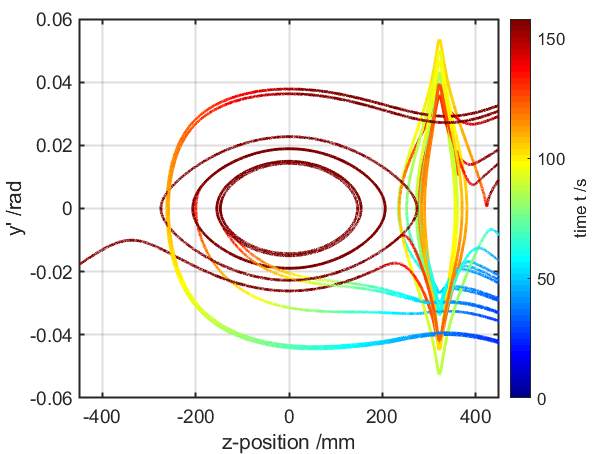}
\label{fig:3DVertPhaseSpace}}
\caption{(a)~Vertical phase-space acceptance of muons injected into the uniform solenoid field. Blue dots indicate the initial vertical position and vertical momentum of all generated muons. Red dots indicate the initial conditions of muons which decay into a positron within the frozen-spin region. (b)~Vertical phase-space plot of muon trajectories. The color code heat map indicates the time after injection. Dark blue to dark red spans a duration of \SI{148}{ns} while brown indicates later times. The magnetic kick is applied at \SI{48}{ns} for a duration of \SI{100}{ns} with half a period of a sinusoidal shape. The peculiar shape at $z\approx\SI{320}{mm}$ coincides with the maximum of the kick amplitude in Figure~\ref{fig:3DBKickField}. Apparently, some muons are reflected by the pulse.}%
\label{fig:3DPhaseSpace}%
\end{figure}

In this first simulation attempt, about 20\% of all muons were finally stored in the central part of the frozen-spin region and decayed to positrons. Figure~\ref{fig:3DVertAcceptance} shows the vertical phase-space acceptance for injection in the injection zone, while Figure~\ref{fig:3DVertPhaseSpace} shows the vertical phase-space trajectories of muons that are stored, closed circles, bypass the central zone, or are reflected. 
For this first simulation we have chosen a field index of $n=2.6E-4$, a further optimisation will need to balance potential systematic effects due to a vertical betatron oscillation and the increase injection efficiency due to a larger vertical phase space. A coupling of the vertical and horizontal phase space prior injection will significantly reduce the vertical divergence in the central plane and hence further improve the injection efficiency~\cite{Iinuma2016NIMA,Rehman2020}. 

\newpage

\subsubsection{Continuous drift measurement} 
\begin{figure}%
\centering
\subfloat[][]{
\includegraphics[width=0.47\columnwidth]{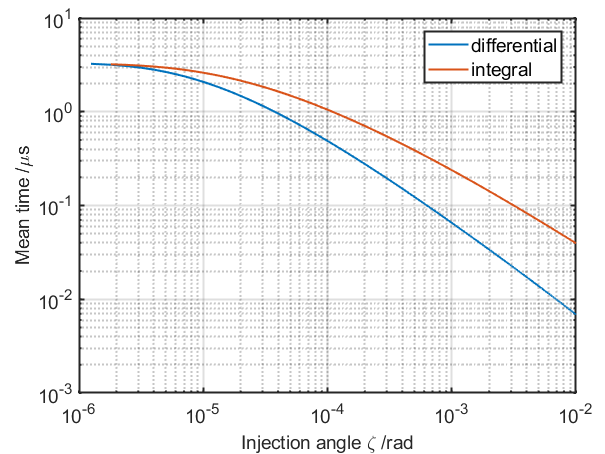}\label{fig:MeanLifeTimeHelix}}%
\hfill
\subfloat[][]{
\includegraphics[width=0.47\columnwidth]{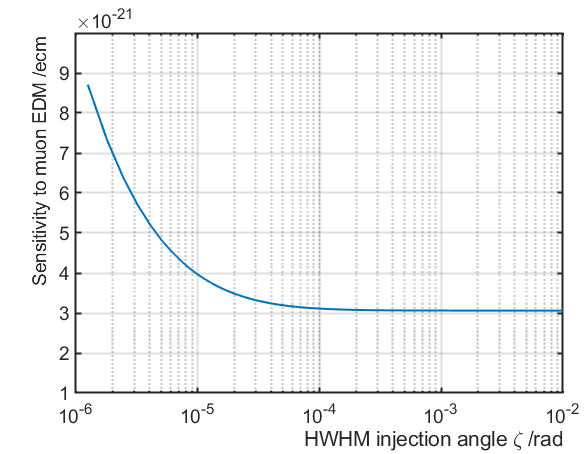}\label{fig:SensitivityHelix}}%
\caption{(a)~The differential, equation~\eqref{eq:MeanTimeMuonHelixMuon}, and integral mean lifetime, equation~\eqref{eq:ZetaAverageMuonTime}, of a muon within the frozen-spin region. (b)~Expected annual sensitivity of the continuous helix muon EDM search connect to the beam line $\mu$E1 as a function of width of the vertical injection angle $\zeta_{\rm inj}$. }%
\label{fig:HelixMuEDM_LifetimeANDSensisitivity}%
\end{figure}

As alternative to a single muon storage ring we also investigated the concept of a continuous injection without triggered magnetic field perturbations. In this case all muons which have passed through the injection channel enter the detector system continuously. A combination of accelerator radio frequency, muon entrance tagger, and positron tracker permits the reconstruction of each event.
The muons will drift parallel to the magnetic field through the frozen-spin field region with $v_\parallel = \beta c \sin(\zeta)$ as function of the effective injection angle $\zeta$.
The average time of a muon with velocity $v_\parallel(\zeta)$ within the central frozen-spin region of length $l$ is

\begin{equation}
 \langle t(\zeta)\rangle = \frac{\int_0^{l/v_\parallel(\zeta)} t \exp\left(-t/(\gamma\tau)\right)}{\int_0^\infty\exp\left(-t/(\gamma\tau)\right)}.
\label{eq:MeanTimeMuonHelixMuon}
\end{equation} 
By averaging over the drift angle $\zeta=\arccos\left(\sqrt{(B_{\rm inj}/B_{\rm c})\cos(\zeta_{\rm inj})^2}\right)$, a smooth function of the injection angle $\zeta_{\rm inj}$ and the magnetic fields $B_{\rm inj}$ and $B_{\rm c}$, one obtains the average time,

\begin{equation}
	\overline{\langle t\rangle} = \frac{\int_0^\zeta \langle t(\zeta)\rangle \mathcal{F}(\zeta) }{\int_0^\zeta \mathcal{F}(\zeta)},
\label{eq:ZetaAverageMuonTime}
\end{equation}
a muon is within the frozen-spin region. 
Here $\mathcal{F}(\zeta)$ is the vertical divergence of the injected beam, typically modelled using a Gaussian distribution. Figure~\ref{fig:MeanLifeTimeHelix} shows the mean-passage times in the case of the measured vertical phase space, see \autoref{sec:muE1}, and a central-field region of $l=\SI{1}{m}$. 
Figure~\ref{fig:SensitivityHelix} shows the expected annual sensitivity for an experiment coupled to $\mu$E1 as a function of width of the distribution of the injection angle $\zeta_{\rm inj}$ and a magnetic-field ratio of $B_{\rm inj}/B_{\rm c}=0.99$.\\

The T-BMT equation is slightly modified in the presence of a constant drift velocity $v_{\theta}$ along the principal magnetic-field direction.
For the sake of calculations, we will use a magnetic field of \SI{3}{T}, muons with a momentum of $125$~MeV/$c$ ($\gamma =1.57$, $\beta=0.77$), and a homogeneous central-field region of length $l=\SI{1}{m}$.
This results in a radius of about $r=\SI{0.14}{m}$ and $v_{\theta} = \SI{1.8e8}{m/s}$. We will use a radial electric field %$E_r=B_z/v_\theta\left(1-1/ac^2/(\gamma^2-1)\right)^{-1}$
$E_r=aB_z/v_\theta\left(1/c^2\left(a-1/(\gamma^2-1)\right)\right)^{-1}$ to establish the frozen-spin condition.
In the case that $v_\parallel=v_z\neq 0$ equation~\eqref{eq:omegaMu1} changes to
%\begin{equation}
%\bvec{\omega}_a = -q/m\left[a
%	\begin{pmatrix}
%		B_\theta \\
%		B_r \\
%		B_z
%	\end{pmatrix}
%	-
%	\frac{a\gamma(v_\theta B_\theta + v_zB_z)}{\left(\gamma+1\right) c^2}
%	\begin{pmatrix}
%			v_\theta \\
%			0 \\
%			-v_z
%	\end{pmatrix}
%	+\left(a-\frac{1}{c^2(\gamma^2-1)}\right)
%	\begin{pmatrix}
%		 v_z E_r \\
%		0 \\
%		v_\theta E_r
%	\end{pmatrix}
%	\right].
%\label{eq:TMBTwithDrift1}
%\end{equation}
\begin{equation}
\bvec{\omega}_a = q/m\left[a
	\begin{pmatrix}
	    B_r \\
		B_\theta \\
		B_z
	\end{pmatrix}
	-
	\frac{a\gamma(v_\theta B_\theta + v_zB_z)}{\left(\gamma+1\right) c^2}
	\begin{pmatrix}
	        0 \\
			v_\theta \\
			v_z
	\end{pmatrix}
	+\frac{1}{c^2}\left(a-\frac{1}{(\gamma^2-1)}\right)
	\begin{pmatrix}
	        0 \\
		    v_z E_r \\
		    -v_\theta E_r
	\end{pmatrix}
	\right].
\label{eq:TMBTwithDrift1}
\end{equation}
In the case of applying the frozen-spin electric field in combination with a uniform magnetic field along the solenoid axis with strength $|\vec{B}|$ and assuming $B_r=0$, we get
%\begin{align}
%\bvec{\omega}_a & =   -q/m\left[a
%	\begin{pmatrix}
%		0 \\
%		0 \\
%		B_z
%	\end{pmatrix}
%	+
%	\frac{a\gamma(v_z B_z)}{\left(\gamma+1\right) c^2}
%	\begin{pmatrix}
%			v_\theta \\
%			0 \\
%			v_z
%	\end{pmatrix}
%	+\left(a-\frac{1}{c^2(\gamma^2-1)}\right)
%	\begin{pmatrix}
%		 -v_z E_r \\
%		0 \\
%		v_\theta E_r
%	\end{pmatrix}
%	\right] \\
%	 & =   -q/m\left[a
%	\begin{pmatrix}
%		0 \\
%		0 \\
%		B_z
%	\end{pmatrix}
%	+
%	\frac{a\gamma(v_z B_z)}{\left(\gamma+1\right) c^2}
%	\begin{pmatrix}
%			v_\theta \\
%			0 \\
%			v_z
%	\end{pmatrix}
%	+a
%	\begin{pmatrix}
%		 v_z/v_\theta B_z \\
%		0 \\
%		-B_z
%	\end{pmatrix}
%	\right] \\
%	& =  -q/m\left[a
%	\frac{\gamma\zeta \beta^2 B_z}{\left(\gamma+1\right)}
%	\begin{pmatrix}
%			\tfrac{2\gamma-1}{\gamma} \\
%			0 \\
%			\zeta
%	\end{pmatrix}
%	\right],
%\label{eq:TMBTwithDrift2}
%\end{align}
\begin{align}
\bvec{\omega}_a & =   q/m\left[a
	\begin{pmatrix}
		0 \\
		0 \\
		B_z
	\end{pmatrix}
	-
	\frac{a\gamma(v_z B_z)}{\left(\gamma+1\right) c^2}
	\begin{pmatrix}
	        0 \\
			v_\theta \\
			v_z
	\end{pmatrix}
	+\frac{1}{c^2}\left(a-\frac{1}{(\gamma^2-1)}\right)
	\begin{pmatrix}
	        0 \\
		    v_z E_r \\
		    -v_\theta E_r
	\end{pmatrix}
	\right] \\
	 & =   q/m\left[a
	\begin{pmatrix}
		0 \\
		0 \\
		B_z
	\end{pmatrix}
	-
	\frac{a\gamma(v_z B_z)}{\left(\gamma+1\right) c^2}
	\begin{pmatrix}
	        0 \\
			v_\theta \\
			v_z
	\end{pmatrix}
	+a
	\begin{pmatrix}
	     0 \\
		 v_z/v_\theta B_z \\
		-B_z
	\end{pmatrix}
	\right] \\
	& =  -q/m\left[a
	\frac{\gamma\zeta \beta^2 B_z}{\left(\gamma+1\right)}
	\begin{pmatrix}
	        0 \\
			\tfrac{1}{1-\gamma} \\
			\zeta
	\end{pmatrix}
	\right],
\label{eq:TMBTwithDrift2}
\end{align}
where $\zeta = v_z/v_\theta$ is the drift angle in the central part of the solenoid. In the next section we show that the continuous drift with  $\zeta<\SI{55}{mrad}$ does not generate a systematic effect larger than $d_\mu \leq  \SI{1e-23}{\ecm}$. However, as Figure~\ref{fig:SensitivityHelix} shows, the continuous muon helix concept is less sensitive than both storage concepts and will only be of interest in an initial phase, if the vertical magnetic kicker is not yet implemented.

\subsection{Systematic effects}
\label{sec:sysEffects}
A excellent starting point for a discussion of systematic effects is provided by the seminal publication by Farley and colleagues\,\cite{Farley2004PRL}. Any non-uniformity or misalignment of the magnetic and electric field and the positron detection system might cause a spin precession or appear as one. As rotations do not commute, particular care has to be taken if several of these effects are combined.

\begin{itemize}
	\item Radial magnetic fields $B_r$
	\item Azimuthal magnetic field $B_\theta$
	\item Vertical electric field $E_V$, i.e.\ $\vec{E}\cdot\vec{B}\neq0$ on orbit
	\item Misalignment in positron detector
	\item Early to late change in detector response	
\end{itemize}

The muons are only stored on orbits where  $\langle B_r \rangle = 0$, where $\langle\,.\,\rangle$ denotes the orbit average. In the cases where the orbit average of the magnetic and electric field components are zero, $\langle B_\theta\rangle =0$ and $\langle E_V\rangle =0$, no systematic effect occur without a remanent $(g-2)$ precession which would lead to non-commutative rotations of the spin. An exact specification to which precision the electric field for the frozen-spin technique needs to be controlled will be derived and cross checked by simulation. This in turn will indicate the required precision for a measurement of the anomalous precession as a function of the applied electric field. The precise knowledge of $a_\mu$\,\cite{Bennett2006PRD} permits us to measure $\langle B\rangle$ on the orbit for $E=0$.

Most effects and combination of effects will cancel when combining clockwise and counter clockwise injection of muons into the spectrometer and averaging data over multiples of orbit periods, $T<\SI{10}{ns}$. A simulation, supporting analytical derivations will specify to which degree the inverse magnetic field for counter clockwise injection needs to be identical to the field for clockwise injection, but in sign.

A specific systematic effect may occur in the less sensitive case were the muons continuously drift through the central part of the spectrometer.
In order to make sure that the drift does not generate a relevant systematic effect the vertical precession $\omega_\zeta$ must be kept much smaller than the sensitivity to an EDM of $d_\mu \leq  \SI{1e-23}{\ecm}$.
Hence,

\begin{align}
	|\omega_\zeta| = \frac{aq}{m}
	\frac{\gamma\zeta^2 \beta^2 B_z}{\left(\gamma+1\right)} & \leq \frac{2d_\mu E^{\ast}}{\hbar} \\
	\zeta^2 & \leq \frac{2m}{aq} \frac{\gamma}{(\gamma-1)B_z} \frac{d_\mu E^{\ast}}{\hbar} \\
%	\zeta^2 & \leq \frac{2m}{aq} \frac{d_\mu \gamma^2 \beta c}{\hbar(\gamma-1)}\\
	\zeta   & \leq \sqrt{2\frac{m d_\mu c}{aq\hbar}} \sqrt{ \frac{\gamma \sqrt{\gamma^2-1}}{(\gamma-1)}}\\
	\zeta & \leq \SI{55}{mrad}\sqrt{\frac{d_\mu}{\SI{1E-23}{\ecm}}}
	\label{eq:etaSystematic}
\end{align}
where the minimum is reached for $\gamma =1.62$, corresponding to $p=\SI{135}{MeV}/c$.
%as visible in Figure~\ref{fig:ZetaVsGamma}.
%

In turn this indicates that any drift angle in the central plane up to \SI{55}{mrad} is acceptable. However, as Figure~\ref{fig:InjectionAndDrift} illustrates the larger the ratio between the magnetic field in the central plane $B_{\rm c}$ and the injection plane $B_{\rm inj}$, the stronger the drift-angle distribution will spread in the central plane.

\subsection{\texorpdfstring{$\mu$E1}{muE1} beam line revisited}
\label{sec:muE1}
The precision measurement of the muon EDM  requires a high-flux polarized muon beam with a small beam emittance as the sensitivity scales with $\sqrt{N}$ and $P_0$. 
%According to equation~(\ref{eq:EDMsensitivity}), such experiment also favours low $\gamma$ values. 
As $E_{\rm f}$ is proportional to $\gamma^2$, equation~(\ref{eq:EDMsensitivity}) indicates that the sensitivity increases with higher $\gamma$ values.
Therefore, this demands the use of a fairly high muon momentum and the $\mu$E1 beam line at PSI is considered to be a potential beam line to host the muon EDM experiment. Note that even higher momenta would result in higher values of the laboratory electric fields needed for the frozen-spin condition, which are more difficult to realise.

\begin{figure}%
	\includegraphics[width=0.6\columnwidth]{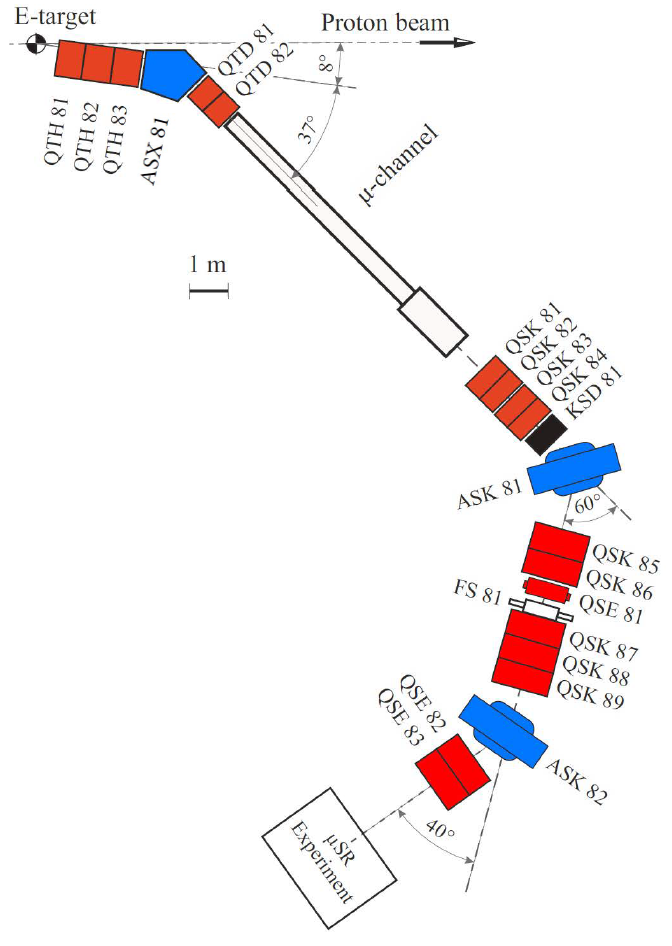}%
	\caption{Layout of the $\mu$E1 area (see \cite{muE1}).}
	\label{fig:muE1_layout}%
\end{figure}

Figure~\ref{fig:muE1_layout} shows the layout of the $\mu$E1 area at PSI: pions produced at target E are extracted, selected in momentum by the dipole magnet ASX~$81$, and then transported through a $5$~T superconducting solenoid, where muons are collected from pion decays followed by the selection of backward decay muons by a second momentum selection performed by the dipole magnet ASK~$81$.

In 2019, a characterization of the $\mu$E1 beam line was performed and the muon-beam rate, transverse phase space (emittance), and polarization level were studied up to the muon-beam momentum of $125$~MeV/$c$ with two different beam line settings, the so-called `new tune' and `$\mu$SR-tune'. Note, that for both settings all configurations of the proton accelerator up to the $\mu$-channel in Figure~\ref{fig:muE1_layout}, were not changed. 

A scintillating fiber (SciFi) beam monitoring detector mounted \SI{526}{mm} downstream of the quadrupole QSE83 (see Figure~\ref{fig:muE1_layout}),was used to measure the muon-beam rate and transverse beam size. Then the transverse phase space was explored by employing a quadrupole-scan technique, which uses the quadratic relationship between the magnetic-field strength of the final focusing quadrupole in the beam line upstream of the beam monitoring detector and the transverse beam size to extract the phase-space parameters, namely Twiss parameters and emittance. Note that such a technique relies on the independent knowledge of the dispersion function for each strength value of the quadrupole used for the scan in order to disentangle the betatronic from the dispersive part of the measured beam size.

\begin{figure}%
 	\includegraphics[width=0.6\columnwidth]{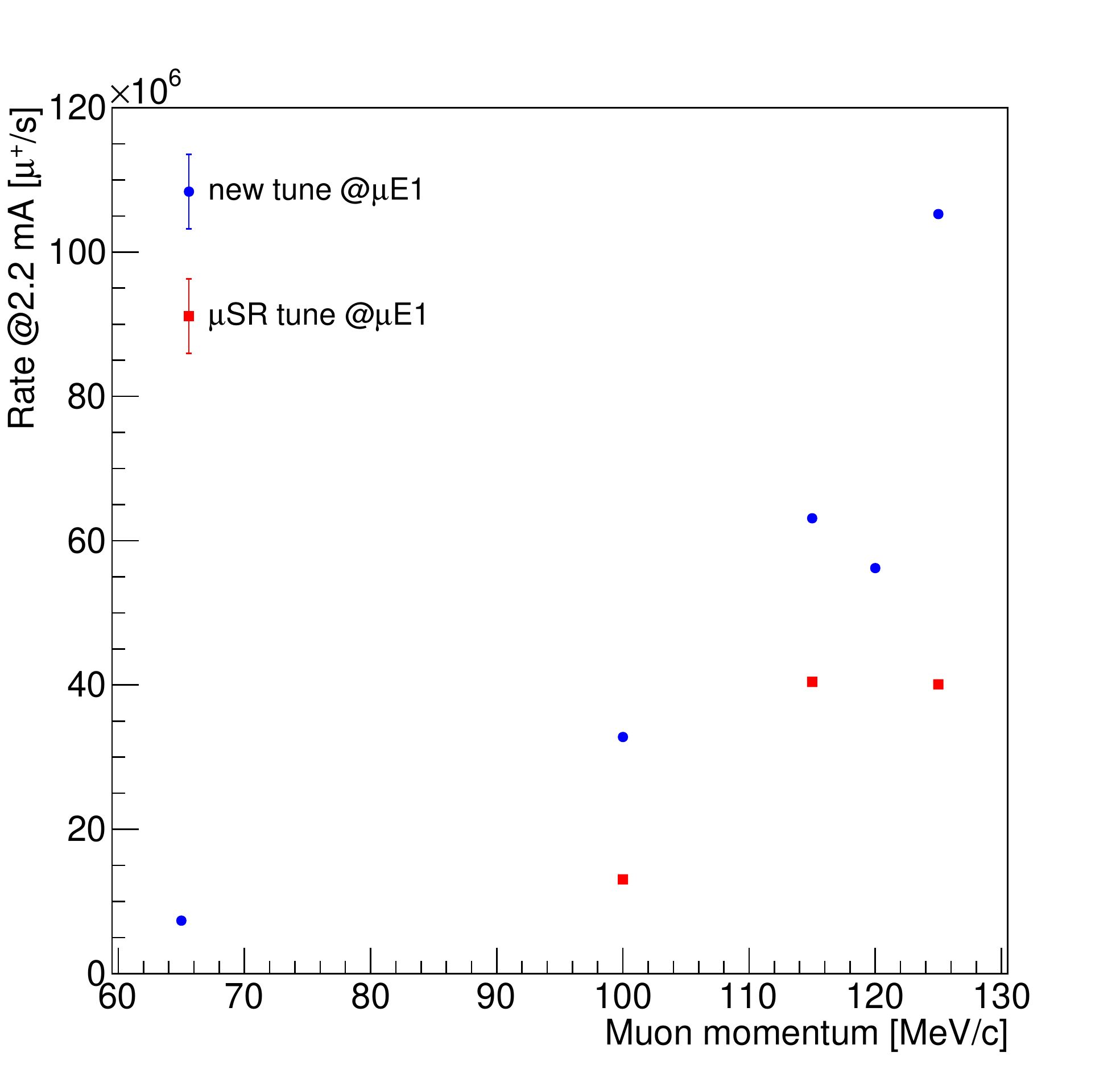}%
 	\caption{Muon-beam rate at the $\mu$E1 beam line for two beam line settings as a function of the muon-beam momentum.}
 	\label{fig:RatevsMom}%
 \end{figure}

\begin{figure}
 	\begin{center}
 		\begin{tabular}{cc}
 			% 1
 			\subfloat[Horizontal]{
 				\begin{minipage}{0.5\hsize}
 					\begin{center}
 						\includegraphics[width=1\columnwidth]{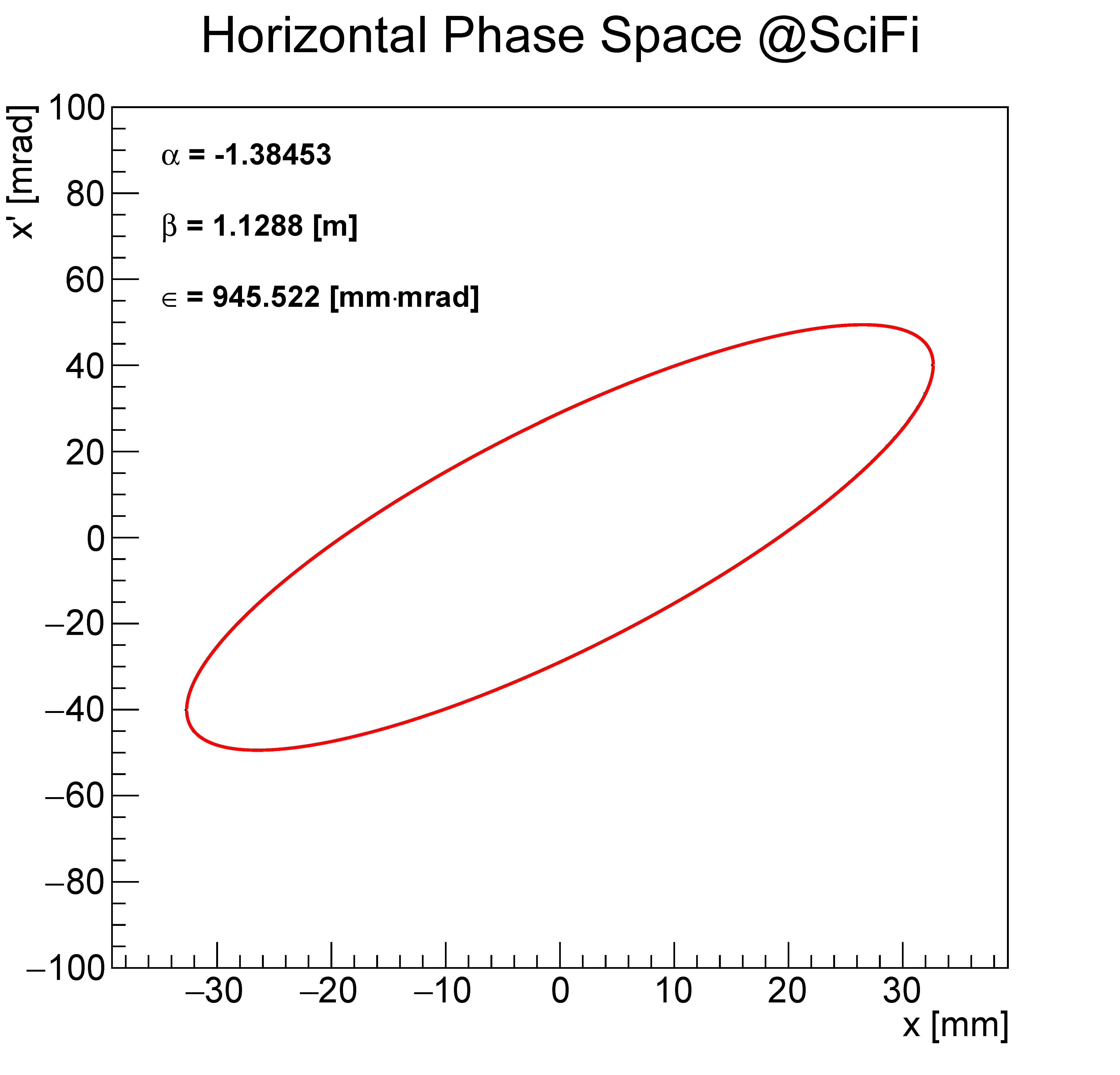}
 					\end{center}
 				\end{minipage}
 			}	
 			% 2
 			\subfloat[Vertical]{
 				\begin{minipage}{0.5\hsize}
 					\begin{center}
 						\includegraphics[width=1\columnwidth]{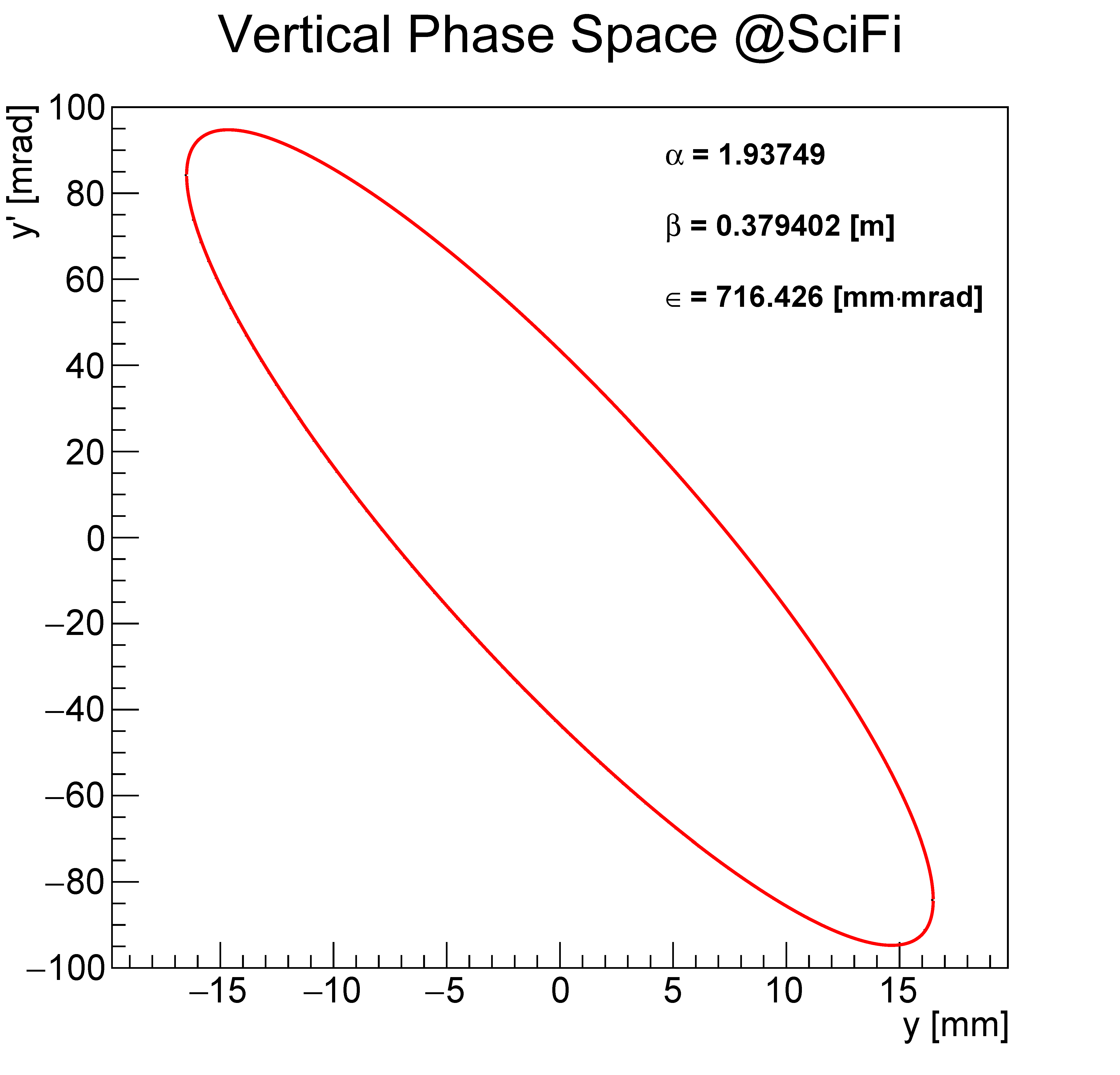}
 					\end{center}
 				\end{minipage}
 			}
 		\end{tabular}
 		\caption{Horizontal~(a) and vertical~(b) phase-space ellipses ($1\,\sigma$) at the SciFi beam monitoring detector position with the new tune at a muon-beam momentum of $125$~MeV/$c$.}
 		\label{fig:PhaseSpace_new_muE1_125MeVc}
 	\end{center}
 \end{figure}

\begin{figure}%
 	\includegraphics[width=0.6\columnwidth]{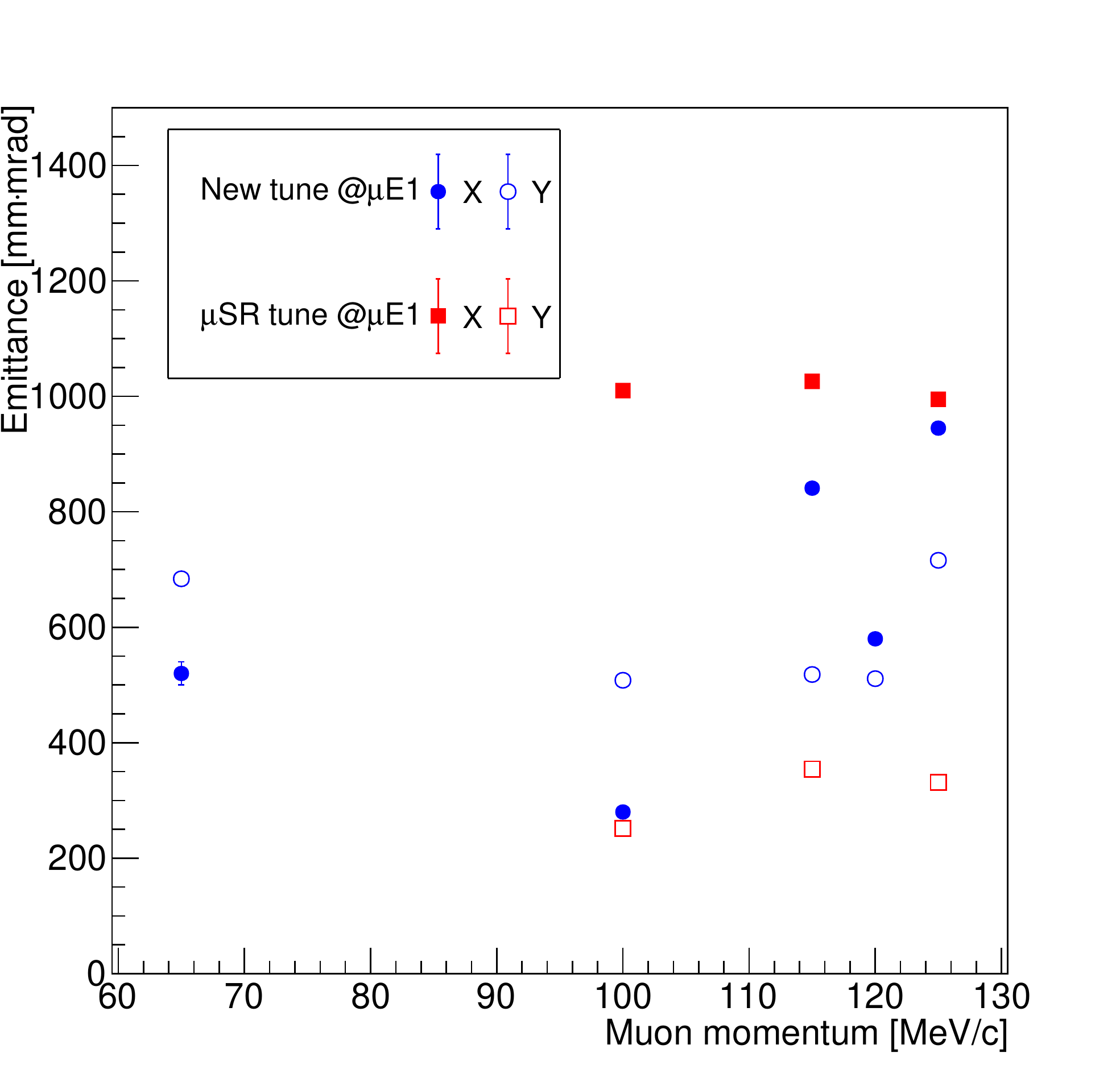}%
 	\caption{Horizontal and vertical emittances ($1\,\sigma$) at $\mu$E1 beam line for two beam line settings as a function of the muon-beam momentum.}
 	\label{fig:EmittancevsMom}%
 \end{figure}

A maximum muon-beam rate of $1.05\times10^{8}$~${\mu}^+$/s at $2.2$~mA proton current was obtained by the new setting of the transfer line, which also minimised the dispersion function at a beam momentum of $125$~MeV/$c$ as shown in Figure~\ref{fig:RatevsMom}. 
Figure~\ref{fig:PhaseSpace_new_muE1_125MeVc} presents the corresponding horizontal and vertical phase-space ellipses with emittances of $945$~$\mathrm{mm}\!\cdot\!\mathrm{mrad}$ and $716$~$\mathrm{mm}\!\cdot\!\mathrm{mrad}$ ($1\,\sigma$), respectively, and Figure~\ref{fig:EmittancevsMom} summarizes the horizontal and vertical emittances for two beam line settings as a function of the muon-beam momentum. 

While we made sure, using \textsc{Transport}-simulations, that for the new beam line setting the beam divergence is large and symmetric around the position `FS81' and close to zero at the position of the beam monitor, this was not the case for the `$\mu$SR-tune' used for solid-state research. Hence, Twiss parameters and emittance for the horizontal planes of the `$\mu$SR-tune' are not completely correct as they still retain a dependence on the dispersion function.
 
 This first characterization serves as starting point to optimize the transfer beam line between the muon decay channel and the muon EDM experiment. If calculations and simulations indicate that an alternative beam layout would further increase the rate and better match the injection phase space, the modification of the beam line shown in Figure~\ref{fig:muE1_layout} is in principle possible. This probably also requires a second test with beam.
 
\begin{figure}%
	\includegraphics[width=0.6\columnwidth]{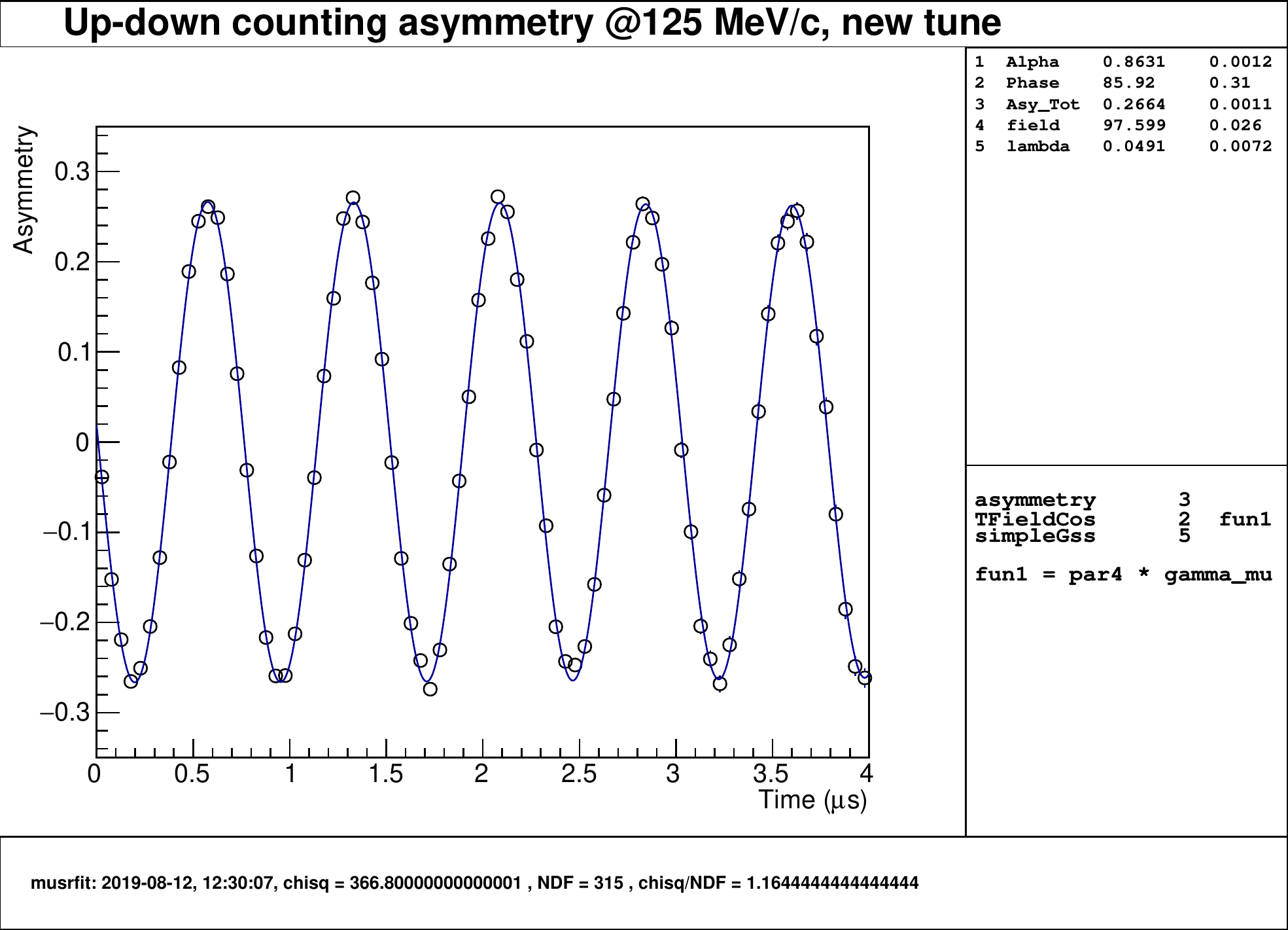}%
	\caption{Measured up-down counting asymmetry plot with the new tune at a muon-beam momentum of $125$~MeV/$c$.}
	\label{fig:Pol_new_muE1_125MeVc}%
\end{figure}

 The polarization measurement was performed with a copper stopping target inside the existing $\mu$SR detector of the GPD instruments and an example of the measured up-down counting asymmetry $A(t)$ at 125~MeV/$c$ with the new tune is shown in Figure~\ref{fig:Pol_new_muE1_125MeVc}:
 \begin{equation}
 	A(t)=\frac{\alpha(N_\uparrow(t)-B_\uparrow)-(N_\downarrow(t)-B_\downarrow)}{\alpha(N_\uparrow(t)-B_\uparrow)+(N_\downarrow(t)-B_\downarrow)},
 \end{equation}
where  $\alpha$ accounts for the different detector efficiencies and solid angles, $N_\uparrow$ and $N_\downarrow$ are the number of positron counts in up and down detectors, respectively, and $B_\uparrow$ and $B_\downarrow$ represent the constant backgrounds in the corresponding detectors. 

\begin{figure}%
	\includegraphics[width=0.6\columnwidth]{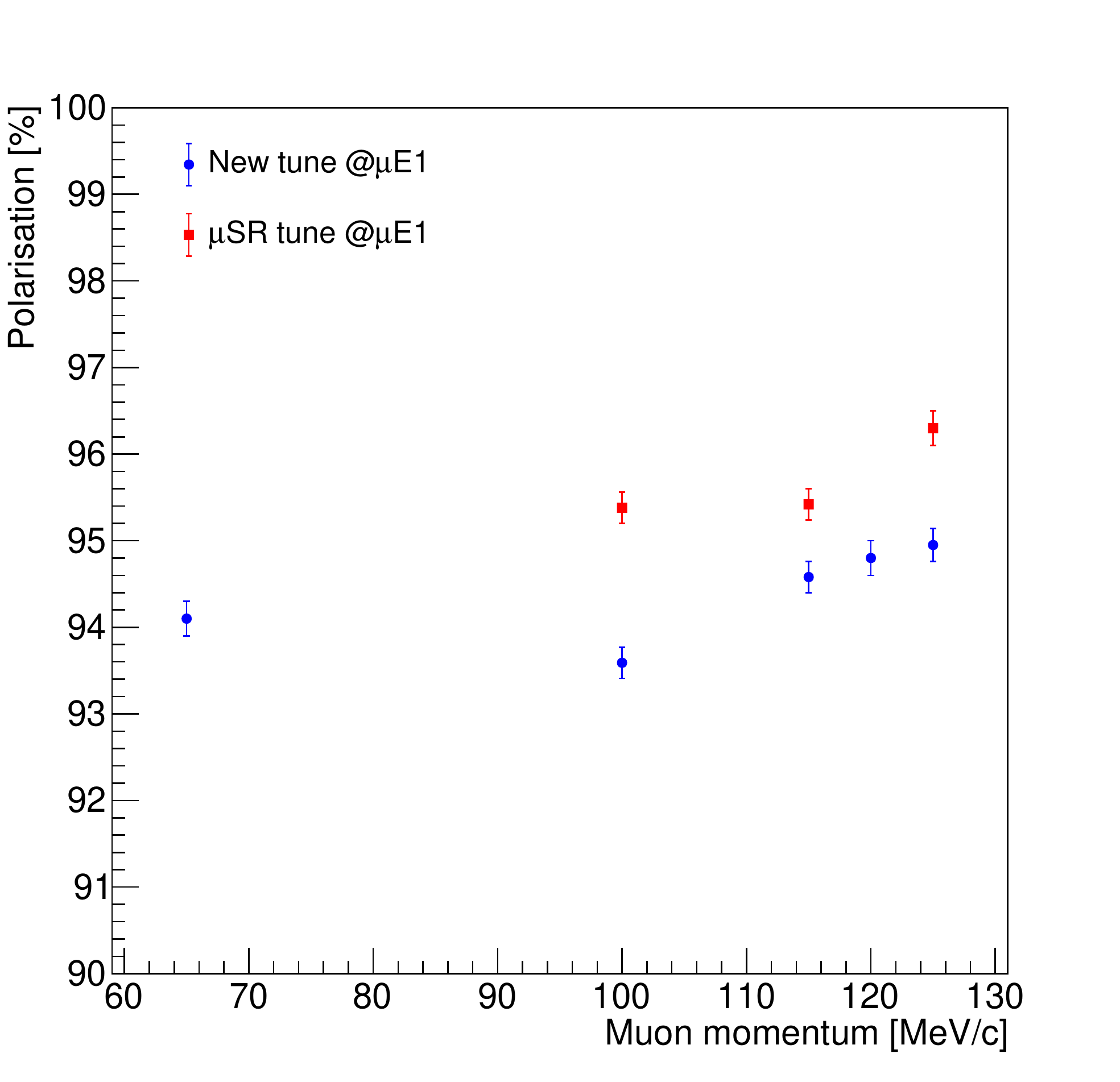}%
	\caption{Initial muon-beam polarization at the $\mu$E1 beam line for two beam tunes as a function of the muon-beam momentum.}
	\label{fig:PolvsMom}%
\end{figure}

Since the oscillation amplitude of $A(t)$ is proportional to the initial muon-beam polarization, the comparison of the amplitude determined from the measurement and the \textsc{Geant4} simulation assuming $100\%$ beam polarization results in an absolute value of the muon-beam polarization. The absolute muon-beam polarization at the $\mu$E1 beam line for both beam tunes as a function of the muon-beam momentum is summarized in Figure~\ref{fig:PolvsMom} and confirms that the initial  polarization is above $93\%$.
\newpage

\subsection{Injection channel}
\label{sec:InjChannel}
All proposed schemes require an injection channel to transport muons from the exit of the beam line through the cryostat, coil package, and vacuum tank into the injection zone inside the magnetic field.
For this purpose, injection using a superconducting magnetic shield, pioneered more than fifty years ago by Firth and coworkers for a \SI{1.75}{T} bubble chamber at CERN~\cite{Firth1970,Firth1973}, and also used for the BNL/FNAL $(g-2)$ experiment~\cite{Yamamoto2002NIMA}, combines many advantages.

The principal idea is that once the superconducting shield is cooled below the critical temperature $T_c$ the field within the injection shield is ``frozen'' even when the outside field is ramped to its nominal strength. Essentially, by ramping the outside field persistent currents will be induced inside the superconductor counteracting to the outside field. This effect is maintained if the shield thickness is sufficiently large for a given outside field and the mean lifetime of the shielding current is long enough. Once the field starts to penetrate, the outside field has to be ramped down and the superconducting shield can be reset by heat cycling.\\

\begin{figure}%
\centering
\subfloat[][]{\includegraphics[width=0.47\columnwidth]{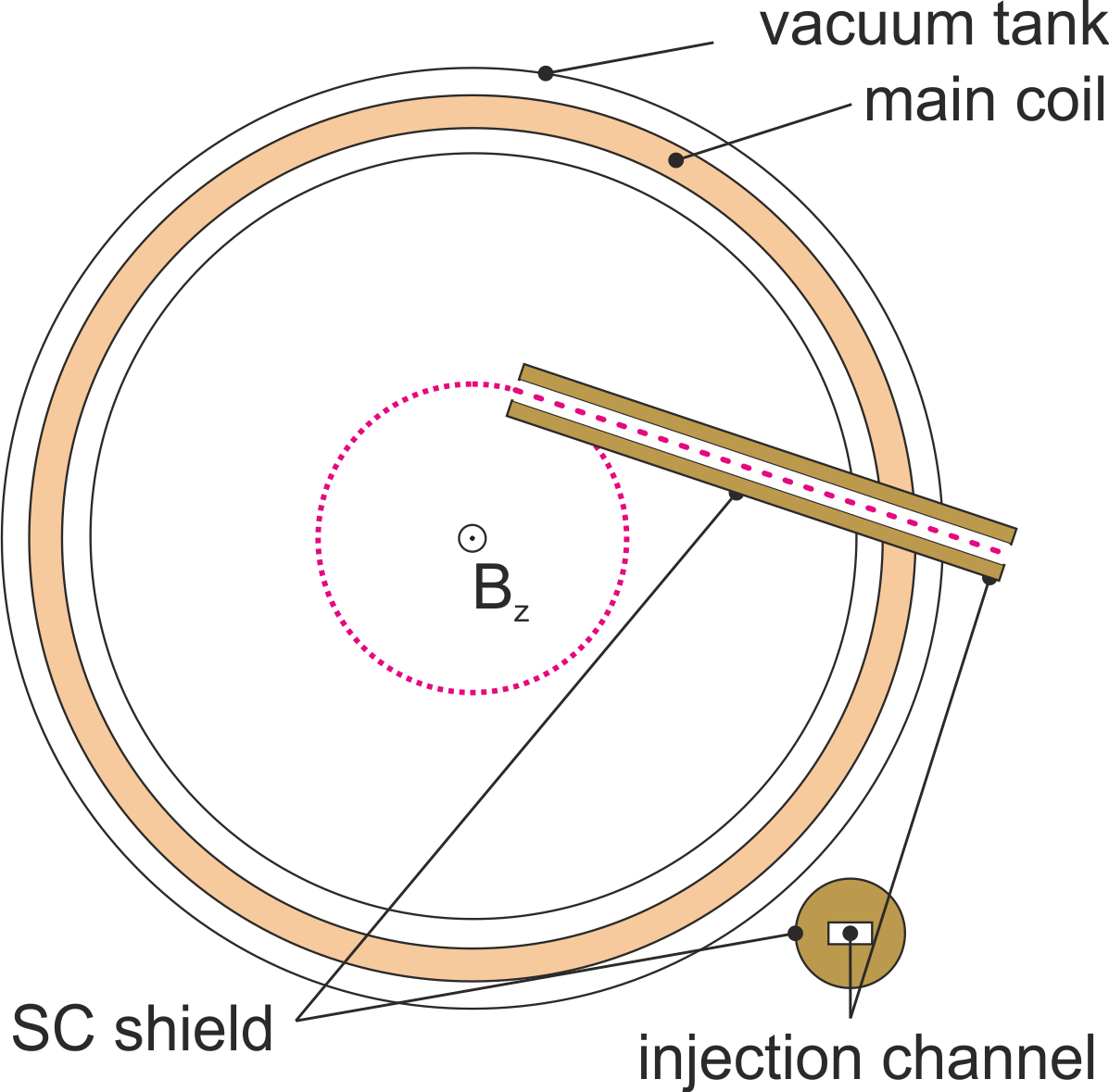}\label{fig:SchemeSCChannel}}%
\hfill
\subfloat[][]{\includegraphics[width=0.47\columnwidth]{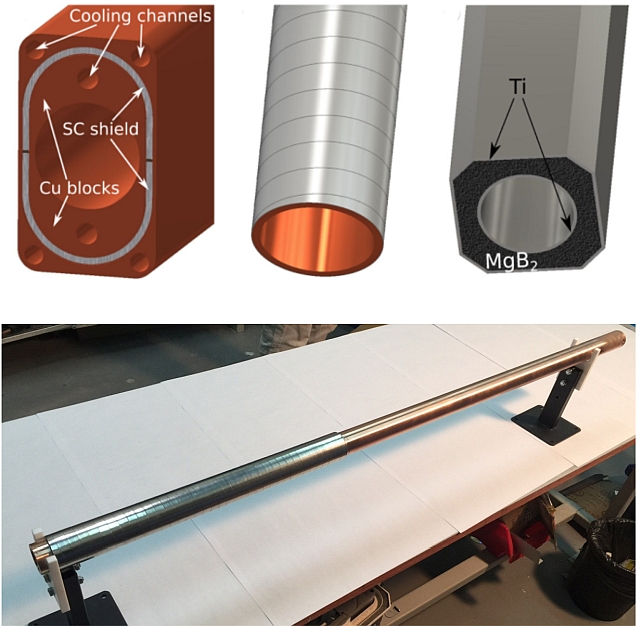}\label{fig:SCOptionsBarna}}%
\caption{(a)~Possible layout for an injection channel based on a shield made of superconductor. For clockwise and counterclockwise injection two superconducting channels will be required. Images~(b) show three different options for the construction of an injection channel. From left to right: multi-layer Nb-Ti/Nb/Cu sheets embedded in a copper block, high-temperature superconducting (HTS) tapes wound in a helical coil onto a copper tube and bulk ${\rm MgB_2}$ sintered into the desired shape. Below an image of the copper tube with HTS superconducting tape. Courtesy of D.~Barna for all images~\cite{Barna2017}. }%
\label{fig:SCChanels}%
\end{figure}

The group of D.~Barna at the Wigner Research Center for Physics in Budapest has investigated three options for a superconducting shield intended for the use in a septum magnet of the CERN FCC~\cite{Barna2017,Barna2018,Barna2019}. Figure~\ref{fig:SCOptionsBarna} on top shows all three variants: multi-layer Nb-Ti/Nb/Cu sheets embedded in a copper block, superconducting HTS tapes wound in a helical coil onto a copper tube and bulk ${\rm MgB_2}$ sintered into the desired shape. Experimental tests demonstrate excellent performance of the multi-layer Nb-Ti/Nb/Cu sheets~\cite{Barna2018} while the sintered variant made of ${\rm MgB_2}$ showed flux jumps at low fields after a successful initial field test~\cite{Barna2019}. The third version using HTS ribbons could not shield the external field, it fully penetrated through the shield above \SI{0.25}{T}. The reason for this might be that the soft-soldering process which was used to attach the ribbons to the copper tube damaged the superconductor. The group of D.~Barna decided not to further investigate the HTS version, but instead started to produce Nb-Ti/Nb/Cu sheets, as the original producer, Nippon Steel Ltd.\ in Tokyo, discontinued production. First sheets will be available by the end of 2020.

For our purpose a HTS version would be favorable as this would not require liquid helium temperatures. However, using the Nb-Ti/Nb/Cu sheets within a sturdy copper structure could be cooled to about \SI{4}{K} using a pulse tube cooler. This might also be attractive for a thermal reset in the case that the external field starts to penetrate.
 
%\newpage
\subsection{Muon identification and trigger}
The concepts to use a magnetic field pulse to kick the muons onto a stable orbit for storage requires an entrance muon detector to trigger the magnetic field. Further, we consider a muon tagger, providing information on the trajectory of the muon, as helpful for a reconstruction of the decay vertex.

\subsubsection{Entrance trigger}
\begin{figure}%
\centering
\includegraphics[width=0.7\columnwidth]{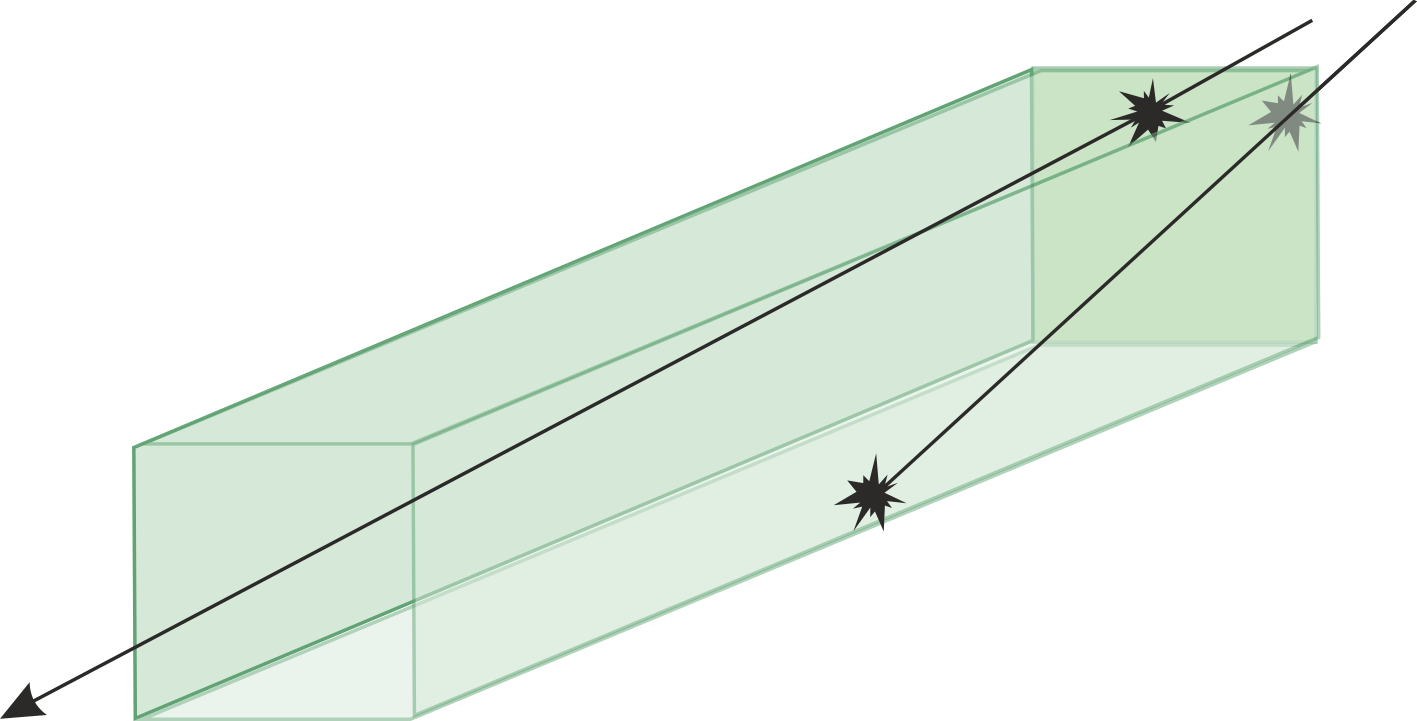}%
\caption{Sketch of the scintillator inside the injection channel, not to scale. Muons enter from the far side, if they pass the channel with out hitting the walls, the magnetic field kick is triggered. Possible dimensions are a cross section of $1\times\SI{1}{cm^2}$ and a length of \SI{100}{cm}.}%
\label{fig:EntranceTrigger}%
\end{figure}

A quick and reliable detection method for an incident muon within the acceptance phase space is required to trigger the magnetic field pulse. A possible scenario is depicted in Figure~\ref{fig:EntranceTrigger}, combining a thin scintillator at the entrance of the injection channel and scintillators on the wall. A robust trigger can be  made by constructing an anti-coincidence between wall scintillators and entrance trigger and including the proton accelerator frequency of \SI{50}{MHz}. In this situation only multiple scattering can occur at the entrance of the injection channel and is controlled by the anti-coincidence. Note, that in average we do not expect more than one muon per accelerator pulse of \SI{20}{ns}.

\subsubsection{Muon tagger}
\label{sec:tagger}
The muon tagger will provide a measurement of time and trajectory of the muon at the injection in the magnetic field. Combined with the time and trajectory of the positron in the central tracker, it will provide the time of flight and positron emission angle to be used in the spin precession analysis. It requires a time resolution below 10~ns and a resolution on the vertical angle of the muon of $\mathcal{O}(\SI{1}{mrad})$. For $\SI{125}{MeV}/c$ muons the multiple Coulomb scattering~(MS) in the detector materials is likely to dominate the tracking resolution of the detector. Moreover, if muons are excessively deflected from their nominal trajectory, the injection efficiency would significantly drop. It means that, in order to provide useful track information and not to compromise the injection efficiency, the material budget of the muon tagger needs to be kept to an extremely low level. For the case of using a different beam-line than $\mu$E1, the muon tagger should also be able to identify electrons and pions possibly contaminating the muon beam.

These requirements favor the use of gaseous detectors with a very light, helium-based gas mixture like helium/isobutane in 90/10 concentration, combined with a thin and possibly segmented fast scintillator in front of the tagger, although some gaseous detectors could also provide timing with the required resolution.

We envisaged a couple of possible configurations for a muon tagger based on gaseous detectors, but we also made a comparison of performances with a design based on solid state detectors and scintillating fibers, that could have the advantage of sharing the detector technology with the positron tracker. 
The guiding principle is to have all detection elements inside the main magnet just after exit of the injection channel.
The sensitive elements of the detector should be distributed along the circular muon trajectory and cover a vertical extent that, depending on the nominal injection angle, can go from $\mathcal{O}(\mathrm{few~cm})$ to $\mathcal{O}(\mathrm{few~10~cm})$. For the determination of the injection angle, the best resolutions should be on the ($\phi$,$Z$) plane. 

\begin{figure}%
\centering
\includegraphics[width=0.7\textwidth]{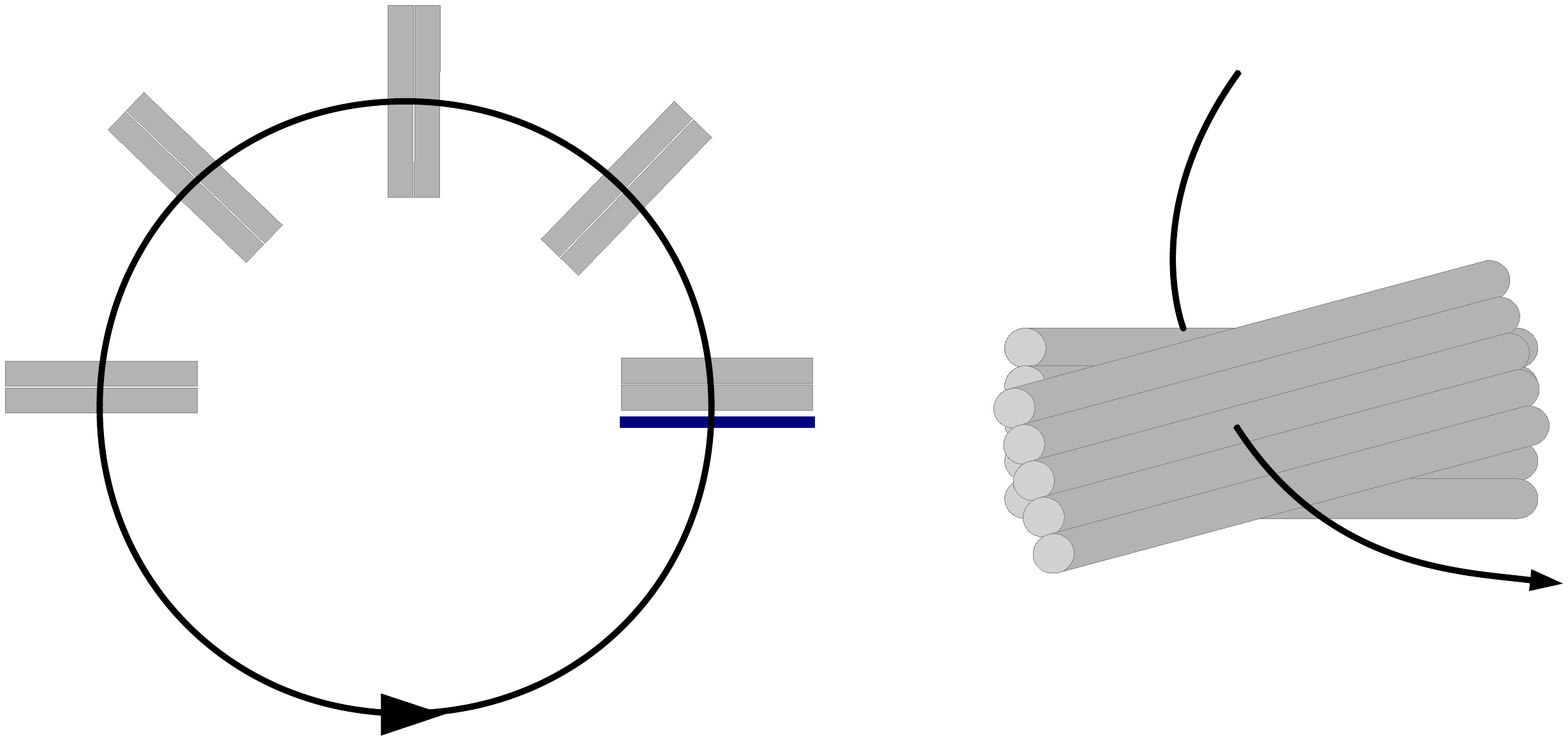}
\vspace{-1cm}
\caption{(Left) Top view of a straw tube tracker with five detector stations (gray) and a scintillator (dark blue) in front of the first station. The trajectory of a positive muon (solid line) within the magnetic field is also shown. (Right) a sketch of a detector station
made of two arrays of straw tubes with a small 
vertical angle among them.}%
\label{fig:straws}%
\end{figure}

If gaseous wire detectors are considered, it implies that radial wires should be used. It makes a traditional, cylindrical drift chamber with longitudinal wires unsuitable. While one could adopt planar drift chambers with very light cathode walls (a small version of the MEG drift chambers~\cite{Adam:2013vqa}, but radially oriented), the necessity of operating the detector in vacuum makes thin-wall straw tubes the most natural choice. They have been already developed to work in vacuum, for instance for the upcoming $\mu \to e$ conversion experiments~\cite{Lee:2016sdb,Nishiguchi:2017gei}, and there is a continuous R\&D effort to further reduce the wall thickness and improve their gas tightness. A set of detector stations, each made of two vertical arrays of radially-oriented tubes, would provide a very good resolution in the ($\phi$,$Z$) plane. With a small vertical angle between the two arrays, each station would also provide a measurement of the radial coordinate. An example of such an arrangement is sketched in Figure~\ref{fig:straws}. In this configuration, the scintillator before the first detector station also provides the track timing that is necessary for the precise determination of the drift time (and hence the drift distance) of ionization electrons in the tubes. As a drawback, straw tubes would give very poor capabilities of separating tracks crossing the detector within $< \SI{1}{cm}$ and $< \SI{1}{\micro s}$.

\begin{figure}%
\centering
\includegraphics[width=0.5\textwidth]{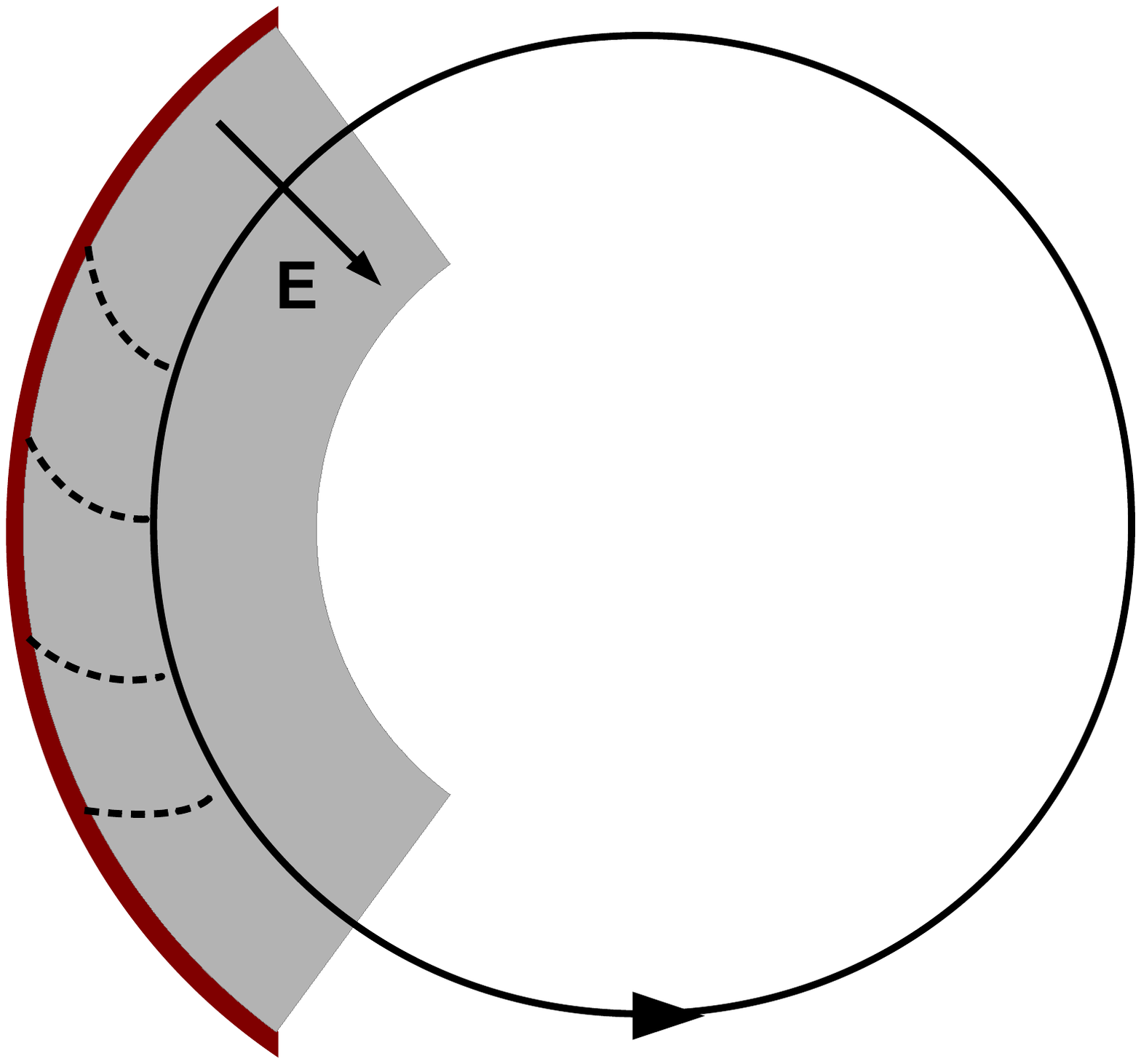}
\caption{Top view of a TPC (gray) with a radial drift
field \emph{E} and GEM readout on the outer surface
(dark red). The trajectory of a positive muon 
(solid line) and the drift lines of ionization
electrons (dashed lines) within the magnetic field 
are also shown.}%
\label{fig:tpc}%
\end{figure}

An alternative configuration could make use of a time projection chamber (TPC) shaped as a cylindrical shell sector (see Figure~\ref{fig:tpc}).
In this case, a radial drift field, with electron amplification and readout placed on the outer cylindrical surface, would provide the required ($\phi$,$Z$) resolution and would reduce
the average drift distance with respect to the usual longitudinal configuration, so to avoid or at least reduce the concentration of heavy additives like CO$_2$ that could be needed to reduce the electron diffusion, in particular when helium is used as a base for the mixture.
Moreover, the drift field could be matched to the electric field used in the spin precession region in order to minimize systematic uncertainties. Cylindrical gas electron multipliers~(GEM)s~\cite{Bencivenni:2007zz} could be used for the amplification and readout. Thin walls will need to be used to contain the gas, but this requirement is limited to the surfaces crossed by the muons. The replacement of traditional GEMs with high-granularity detectors like GEMPix~\cite{George_2015} or GridPix~\cite{vanderGraaf:2007zz} could provide an extremely high capability of resolving multiple tracks crossing the detector within $\Delta t < \SI{1}{\micro s}$, so that the same technology could be used for future upgrades of the experiment, when pileup could become an issue. Also for a TPC, a scintillator has to provide the starting time for the drift distance measurement.

If solid state detectors or scintillating fibers are considered, radial planes of such devices should be used, with one layer of pixels or two crossed layers of fibers. With currently available technologies, the minimum thickness of silicon detectors and scintillating fibers that could suit this application are \SI{50}{\micro m} and \SI{250}{\micro m}, respectively.

\begin{table}[h]
    \centering
    \begin{tabular}{|l|p{4cm}|c|c|}
    \hline
    \hline
    Technology & Detector materials & Single-hit resolution & number of 3D hits\\
    \hline
    TPC & He:i-C$_4$H$_{10}$ (90:10) \newline 25~$\mu$m Mylar\textsuperscript{\tiny\textregistered} walls & 300~$\mu$m & 3.3 per cm\\    
    \hline
    Straw tubes & He:i-C$_4$H$_{10}$ (90:10) \newline 12.5~$\mu$m Mylar\textsuperscript{\tiny\textregistered} tubes & 100~$\mu$m & 1 per station (2 tubes) \\
    \hline
    Silicon pixels & 50~$\mu$m Silicon & 10~$\mu$m & 1 per station\\    
    \hline
    Scintillating fibers & 250~$\mu$m plastic scintillator & 72~$\mu$m & 1 per station (2 layers)\\    
    \hline
    \hline
    \end{tabular}
    \caption{Overview of different technologies for the muon tagger.}
    \label{tab:muon_reso}
\end{table}

All these devices would provide a resolution of $\mathcal{O}(\SI{100}{\micro m})$ or better on the vertical position, and the main uncertainty on the muon track extrapolation would come from the resolution on the vertical angle. 
We performed calculations to infer the resolution that could be reached by detectors built with the aforementioned technologies, under the assumptions summarized in Table~\ref{tab:muon_reso}. 

\begin{figure}%
\centering
\includegraphics[width=0.9\textwidth]{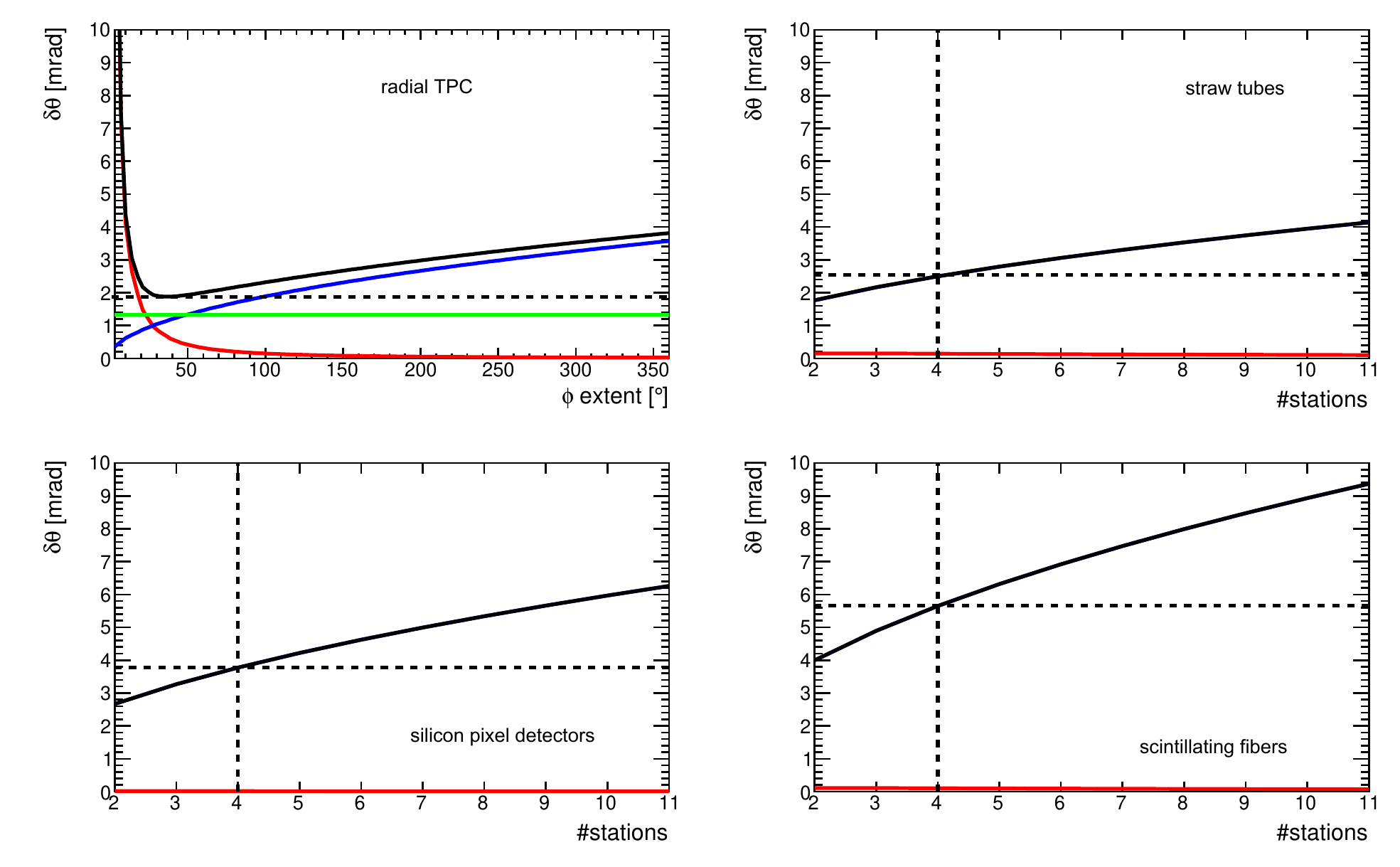}
\caption{Total expected resolution (black) on the vertical angle from a muon tagger made of a radial TPC (top left) or a set of detector stations with straw tubes (top right), silicon pixel detectors (bottom left) or scintillating fibers (bottom right), as a function of the angular extent of the TPC or the number of detector stations. The contributions of the single-hit resolution (red), the MS in the detectors (blue) and the MS in the exit wall of the TPC (green) are also shown. The resolution of straw tubes, silicon pixels and fibers is dominated by the MS and the blue line in hidden behind the black one.}%
\label{fig:muon_reso}%
\end{figure}

We used approximated formulae~\cite{osti_4142694} assuming that a simple $\chi^2$ fit of a straight line to the measured points is performed on the ($\phi$,$Z$) plane in a uniform magnetic field. We considered the contribution of the single-hit position resolution, the multiple Coulomb scattering in the active material of the detector and, for the TPC, the MS in the chamber wall that is crossed to exit the detector. 
It is assumed to be a Kapton\textsuperscript{\tiny\textregistered} or Mylar\textsuperscript{\tiny\textregistered} foil of \SI{25}{\micro m} thickness, although a dedicated R\&D would be required to determine the minimum thickness needed to bear the gas pressure against the vacuum outside the detector. 
The results of our calculations are shown in Figure~\ref{fig:muon_reso} as a function of the $\phi$ extent of the TPC or the number of detector stations for the other technologies, with being four the minimum number of 3D points that is needed to fit a helix. 
As expected, the amount of material in silicon detectors, scintillating fibers, and even straw tubes makes the contribution of the single-hit resolution subleading with respect to the MS. The best performances are obtained with a TPC, with an expected resolution slightly below \SI{2}{mrad}. It is important to notice that, in such a MS-dominated tracking problem, a significant improvement of the resolution with respect to these calculations can be obtained with a broken-curve fit, usually solved with the Kalman-Filter technique~\cite{Fruhwirth:1987fm}. Nonetheless, the MS in the last detector or in the exit wall of the TPC gives an irreducible contribution to the resolution, that is about \SI{1}{mrad} under the assumptions above.

%\newpage

\subsection{Pixel tracker for positrons}

Silicon pixel detectors became a workhorse in many tracking applications. The advent of DMAPS allowed for very thin detectors, suitable for the detection of low-momentum particles like in this application. The Mu3e collaboration developed a pixel detector using this technology~\cite{Arndt:2020obb}. A thickness of $\approx\SI{0.1}{\percent}$ of radiation length has been reached by combining a \SI{50}{\micron} thin silicon pixel chip with an aluminium-polyimide-based high-density interconnect. The chip dimensions are approximately $\SI[parse-numbers = false]{20\times23}{\mm\squared}$ with an active region of about $\SI[parse-numbers = false]{20\times20}{\mm\squared}$ per chip. With some restrictions due to the electrical connections, detector modules with arbitrary shapes can be designed and assembled to cover the desired area. Proven designs for detector barrels exist from Mu3e, made up with modules arranging up to 18~chips in a row. The cooling of the pixel chips will be done using gaseous helium, as in Mu3e. This will require a volume separation between the vacuum and the location of the pixel detector. A similar development is currently in progress at PSI for a prototype study to build a detector for muon spin rotation experiments.
% Results are expected within the next few years.

\begin{figure}%
\centering
    \includegraphics[width=0.5\textwidth]{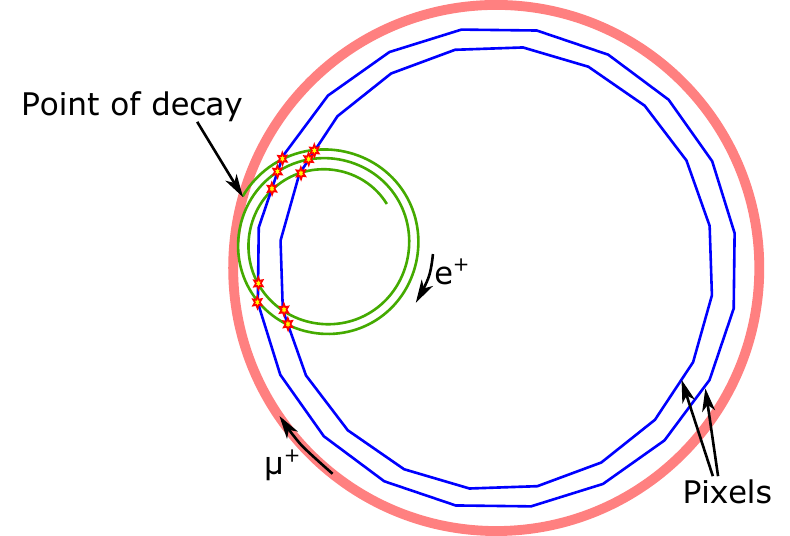}
    \caption{A rough sketch of a possible pixel sensor arrangement. Pixel modules are arranged concentric around the muon trajectories inside the storage ring. Upon a decay, positrons escape and follow a circular trajectory until they get lost. Hits of such a trajectory are marked with asterisk. Note: depending on the choice of magnetic field, an arrangement of pixels inside the muons or on both sides might be chosen.}
    \label{fig:muEDMpixelProposal}
\end{figure}

Figure \ref{fig:muEDMpixelProposal} shows a possible configuration where positrons would be tracked outside of the muon trajectory. The polygonal shapes are two layers of pixel barrels. The indicated example trajectory of a positron demonstrates how the hit information can be reconstructed to determine a)~the decay vertex of the muon, b)~the direction of the positron, and c)~to determine the momentum of it. The momentum gets reduced in every turn due to the material being traversed, and eventually the positron will escape the detector plane. From experience with Mu3e, the expected resolution for the vertex extrapolation would be within $\mathcal{O}(\SI{200}{\micron})$ (dominated by multiple scattering the volume separation) and $\mathcal{O}(\SI{1}{\percent})$ for the momentum. A scintillator based end trigger will supplement the positron tracker to provide for a rapid signal to raise the injection veto. 

\newpage
\section{Future prospects using HIMB and muCool}
\label{sec:HiMBMuCool}
At PSI, the HIMB project is currently being pursued that aims at delivering a surface muon beam of $10^{10}$~$\mu^+$/s to two experimental areas for next-generation particle physics experiments and novel methods to perform muSR measurements. %Figure~\ref{fig:HIMB} shows the overall concept.
The existing \SI{5}{mm} long TgM is replaced by a slanted, \SI{20}{mm} long target TgH that emits around $10^{11}$~surface-$\mu^+$/s to either side. In order to capture a large fraction of this flux, two normal-conducting, radiation-hard solenoids are placed \SI{250}{mm} away from the target. After this initial capture and in order to  reach a high overall efficiency, the use of large-aperture solenoids and dipoles is continued along the beam line. Preliminary simulations showed that in that way around \SI{10}{\%} overall capture and transmission efficiency can be reached leading to surface muon rates of around $10^{10}$~$\mu^+$/s in the experimental areas. The project is currently developing the Conceptual Design Report. The realization is foreseen during the Swiss funding period 2025-2028.

%\begin{figure}%
%\centering
%\includegraphics[width=0.9\columnwidth]{HIMB_concept}%
%\caption{Sketch of the HIMB concept: The existing target %station TgM is replaced by a new target station TgH with %two high efficiency solenoid beamlines delivering %$10^{10}$~$\mu^+$/s to two experimental areas for %particle physics and muSR.}%
%\label{fig:HIMB}%
%\end{figure}
%
In addition to HIMB, the muCool project is also progressing. Within this project the cooling of a positive muon beam is studied in order to reduce its six-dimensional phase space by a factor $10^{10}$. 
%Figure~\ref{fig:muCool} shows the general concept.
A standard muon beam is injected into a cryogenic helium gas cell and stops. The combination of strong magnetic and electric fields together with a position-dependent gas density allows to compress the stopped muon beam both in transverse and longitudinal directions into a point and extract the muons through a small orifice back into vacuum \cite{Taqqu2006}. While both the transverse and longitudinal compression stages have been demonstrated experimentally \cite{Bao2014PRL, Antognini2020PRL}, work is currently ongoing on their combination and the extraction of the muons into vacuum. The extracted muons will be accelerated by a pulsed electrode system to around \SI{10}{keV} energy and taken out of the strong magnetic field by terminating it non-adiabatically. The final low-energy muon beam will have a transverse size of around \SI{1}{mm} and an energy of \SI{10}{keV} with 10~eV spread. Due to the expected efficiency of $10^{-3}$ and the reduced six-dimensional phase space by a factor $10^{10}$, the overall brightness of the beam will increase by seven order of magnitude.
%
%\begin{figure}%
%\centering
%\includegraphics[width=0.7\columnwidth]{muCool_concept}%
%\caption{Drawing of the muCool apparatus used for %compressing the phase space of a standard muon beam by 10 orders of magnitude. For more details see text.}%
%\label{fig:muCool}%
%\end{figure}
%
Within the science case of the HIMB project, re-acceleration of the muCool beam to higher energies than the planned \SI{10}{keV} within the muCool project will be studied. While we currently do not see any reason why such a re-acceleration should not be possible with high efficiency to the momenta required by a muon EDM search, the whole scheme still needs to be worked out. The combination of HIMB, muCool and re-acceleration would deliver a very bright beam of around $10^{7}$~$\mu^+$/s that could be injected into the storage ring or solenoid discussed in the previous sections with high efficiency and would allow to push the EDM search to its limit.

\section{Conclusions}
The search for an electric dipole moment of the muon is a great science opportunity to unveil new sources of CP violation and to test lepton universality in one of the least tested domains of particle physics. The proposed frozen-spin approach in combination with a three-dimensional injection is a novel concept permitting to search for a muon EDM with an unprecedented sensitivity of better than \SI{6e-23}{\ecm}.

%In the next two years the authors will work on a detailed  technical design and ask for funding. We plan to present to the research committee a full experimental proposal by 2022 and a technical design report by January 2023.

\section{Acknowledgments}
We acknowledge the instructive discussions and advice from D.~Barna on superconducting magnetic shielding. The `kick-off' workshop initiating this project was funded by the SNSF No. 191973  grant for scientific exchange. We would like to thank  M.~Hoferichter, M.~Ramsey-Musolf, A.~Signer, and M.~Spira for stimulating discussions of theory aspects.

\input{LoI_MuEDM_bbl}
%\bibliographystyle{NumEtAlTitleDOI}
%\bibliography{nEDM-references,SM-references,MuonReferences,BSM-references,ReferencesIntro,tracker-references}
%
\end{document}

%% file: CommandsAndShortcuts.tex
\newlength{\bildtitel}
\newcommand\REVIEW[1]{\message{LaTeX Warning: \noexpand untreated nEDM-REVIEW command in \jobname .tex: l\the\inputlineno}}% for publication
\newcommand{\bvec}[1]{\ensuremath{\vec{\bm{#1}}}}
\newcommand{\diff}[1]{\operatorname{d}\ifthenelse{\equal{#1}{}}{\,}{\!#1}}
\newcommand{\Diff}[2]{\displaystyle\frac{\diff{#1}}{\diff{#2}}}

\newcommand{\pow}[2]{\ensuremath{#1\!\times\!10^{#2}}}

\newcommand{\ecm}{\ensuremath{\si{\elementarycharge}\!\cdot\!\cm}}

\newcommand{\cm}{\ensuremath{\mathrm{cm}}}
\newcommand{\mm}{\ensuremath{\mathrm{mm}}}